\documentclass[10pt]{iopart}

\usepackage{citesort}
\usepackage{graphicx}% Include figure files
\usepackage[caption=false]{subfig}
\usepackage{amsgen}
\usepackage{amsfonts}
\usepackage{amsbsy}
\usepackage{amssymb}
\usepackage{iopams}
\usepackage{setstack}
\usepackage{bbold}
\usepackage{braket}
\usepackage{xcolor}

\newcommand{\coloneqq}{\mathrel{\mathop:}=}
\newcommand{\inn}[2]{\langle{#1}|{#2}\rangle}

\makeatletter
\renewcommand{\@appendixstar}{\setcounter{section}{0}
 \setcounter{subsection}{0}
 \setcounter{equation}{0}
 \setcounter{figure}{0}
 \setcounter{table}{0}
 \def\thesection{\appendixname\ \Alph{section}}
 \def\theequation{\Alph{section}.\arabic{equation}}}
\renewcommand{\appendixname}{Appendix}
\makeatother

\newtheorem{Thm}{Theorem}

\newtheorem{Prop}[Thm]{Proposition}

\begin{document}

\title{Formulas for Schmidt Decompositions of Mutually Orthogonal Quantum States in Two-Qubit Systems}

\author{Yonghae Lee$^1$, Youngho Min$^2$, Sunghyun Bae$^3$ and Youngrong Lim$^{4,5,\dagger}$}
\footnotetext[0]{ Corresponding author. E-mail: \mailto{sshaep@gmail.com}}
\address{$^1$ Department of Liberal Studies, Kangwon National University, Samcheok 25913, Republic of Korea}
\address{$^2$ Ingenium College of Liberal Arts, Kwangwoon University, Seoul 01897, Republic of Korea}
\address{$^3$ Department of AI convergence Electronic Engineering, Sejong University, 209, Neungdong-ro, Gwangjin-gu, Seoul 05006, Republic of Korea}
\address{$^4$ Department of Physics, Chungbuk National University, Cheongju, Chungbuk 28644, Korea}
\address{$^5$ School of Computational Sciences, Korea Institute for Advanced Study, Seoul 02455, Korea}

\begin{abstract}
We present Schmidt decomposition formulas for mutually orthogonal two-qubit pure states and classify orthonormal sets according to their entanglement structure. We begin by presenting explicit formulas for the Schmidt decomposition of arbitrary two-qubit pure states. Building on this, we construct and analyze all possible configurations of two mutually orthogonal pure states, providing corresponding Schmidt decomposition expressions for each case. For orthonormal sets of three and four states, we derive explicit formulas for selected representative types and discuss the difficulties in obtaining closed-form expressions for more general cases. In particular, we focus on special classes of orthonormal bases that include one or two maximally entangled states. Finally, we prove that no orthonormal basis can contain three product states and one entangled state, revealing a structural constraint that restricts the composition of entangled orthonormal sets.
\end{abstract}

\vspace{2pc}
\submitto{\jpa}
\maketitle

%%%%%%%%%%%%%%%%%%%%%%%%%%%%%%%%%%%%%
%%%
%%% Introduction
%%%
%%%%%%%%%%%%%%%%%%%%%%%%%%%%%%%%%%%%%
\section{Introduction}

Entanglement is a fundamental concept in quantum information processing, enabling tasks such as quantum teleportation~\cite{Bennett1993} and superdense coding~\cite{Bennett1992}. In quantum information theory, it is important to determine whether a given state is entangled and, if so, to quantify its entanglement. For bipartite pure states, the Schmidt decomposition~\cite{Schmidt1907,Peres1995,Nielsen2010,Wilde2013} serves as a key tool for this purpose. Specifically, the Schmidt rank indicates whether the state is a product or entangled, and the von Neumann entropy of a quantum subsystem, which quantifies the entanglement, is derived from the Schmidt coefficients.

Because of the close relation between the Schmidt decomposition and the entanglement of the state, a variety of directions have been explored. For three-qubit systems, explicit Schmidt-type analyses have been presented in~\cite{Acin2000} and have been extended to general tripartite systems in~\cite{Pati2000}. More broadly, multipartite generalizations and structural characterizations have been developed in~\cite{Carteret2000,Huber2013,Kumar2024,Kumar2025}. Beyond a single state, decompositions involving two or more states have been investigated in~\cite{Hiroshima2004,Eltschka2020}, depending on whether several states can be simultaneously decomposed by the same basis. Related variants include operator-level generalizations of Schmidt rank together with corresponding operator decompositions~\cite{Luijk2024,Mansuroglu2024}, and an extension of the notion of Schmidt vectors from pure to mixed states~\cite{Terhal2000,Meroi2024}.  

Motivated by the previous results of the Schmidt decomposition for several states~\cite{Hiroshima2004,Eltschka2020}, we consider the Schmidt decomposition of a set of orthogonal pure bipartite states. Although the original Schmidt decomposition gives Schmidt coefficients for each state, constructing such a set of states is not trivial because of the orthogonality condition between any two states in the set. To the best of our knowledge, such an orthogonality within Schmidt decomposition has not been carried out even in the two-qubit case; our results fill this theoretical gap.

The main contribution of our work is the incorporation of orthogonality into the framework of Schmidt decomposition to enable the classification of mixed states based on their spectral decompositions. Specifically, for a fixed rank, we explicitly construct mixed states whose eigenstates are either product or entangled states, including cases where some of the eigenstates are maximally entangled. These constructions are presented in the form of Schmidt decompositions under orthogonality constraints. While it is, in principle, possible to derive general formulas for all orthogonal configurations, the resulting expressions become prohibitively complex and offer limited practical utility. Instead, our approach provides a systematic method for generating Schmidt decompositions with orthogonality in the two-qubit setting. Notably, we prove that three product states and one entangled state cannot coexist in an orthonormal basis that admits a Schmidt decomposition. This structural constraint connects to the theory of unextendible product bases~\cite{Bennett1999} and has implications for local state discrimination of the bases~\cite{horodecki2003local}.

The potential applications of our results are closely aligned with established uses of the Schmidt decomposition and the Schmidt number. In particular, these notions have been employed in the construction of entanglement witnesses~\cite{Sanpera2001,Wyderka2023,Shi2024,Zhang2024,Li2025}, the certification~\cite{Mukherjee2025} and concentration~\cite{Krebs2024} of entangled states, the analysis of entanglement distribution in pure non-Gaussian tripartite states~\cite{Merdaci2024}, and the characterization of quantum channels~\cite{Mallick2024}. We expect that explicit formulas for the Schmidt decomposition of two-qubit pure states can provide finer analytical control in these settings and may facilitate more precise criteria or performance guarantees.

This paper is organized as follows. Section~\ref{sec:1} presents explicit formulas for calculating the Schmidt decomposition of a given two-qubit pure state. Section~\ref{sec:2} introduces Schmidt decomposition formulas for any two orthogonal states. Section~\ref{sec:3} extends this framework to orthonormal sets of size 3, providing formulas for specific types and addressing the challenges in deriving formulas for the others. In Section~\ref{sec:4}, we demonstrate that an orthonormal basis consisting of three product states and one entangled state cannot exist. We also present Schmidt decomposition formulas for some types of orthonormal bases. Section~\ref{sec:Examples} illustrates the verification of our formulas via representative examples and explains how our results can be used to design state-preparation circuits. Finally, Section~\ref{sec:Conclusion} summarizes our findings. Table~\ref{tab:summary} provides an overview of our main results and their structural relationships.

\begin{table}
\caption{\label{tab:summary}
Overview of our main results. We classify orthonormal sets of two-qubit states according to the number of product (P) and entangled (E) states, and present existence and construction results for each type via Schmidt decompositions. For orthonormal sets containing two or more entangled states, we do not provide full decomposition formulas, as the expressions—though in principle derivable—become prohibitively complex for practical use (dashes in the table). This complexity is discussed in detail in Sec.~\ref{sec:3}(iii). Instead, for PEEE- and EEEE-type orthonormal bases, we present formulas for special cases in which the basis includes one or more maximally entangled (M) states.}
\footnotesize
\begin{tabular}{@{}ll}
\br
\textrm{Types} & \textrm{Remarks} \\
\mr
PP & See Proposition~\ref{prop:PP} \\
PE & E may be diagonal or non-diagonal; see Propositions~\ref{prop:PE:D},~\ref{prop:PE:ND} \\
EP & See Proposition~\ref{prop:EP} \\
EE & The second E may be diagonal or non-diagonal; see Propositions~\ref{prop:EE:D},~\ref{prop:EE:ND} \\
PPP & See Proposition~\ref{prop:PPP} \\
PPE & The second P divides into three distinct cases; see Propositions~\ref{prop:PPE1},~\ref{prop:PPE2},~\ref{prop:PPE3} \\
PEE & ----- \\
EEE & ----- \\
PPPP & See Theorem~\ref{thm:PPPP} \\
PPPE & Non-existence proven; see Theorem~\ref{thm:PPPE} \\
PPEE & The second P divides into three distinct cases; see Theorems~\ref{thm:PPEE1},~\ref{thm:PPEE2},~\ref{thm:PPEE3} \\
PEEE & ----- \\
EEEE & ----- \\
PMEE & Special case of PEEE: see Theorem~\ref{thm:PMEE} \\
MMEE & Special case of EEEE: either both E's are diagonal or both E's are non-diagonal; see Theorems~\ref{thm:MMEE:D},~\ref{thm:MMEE:ND} \\
\br
\end{tabular}
\end{table}

%%%%%%%%%%%%%%%%%%%%%%%%%%%%%%%%%%%%%%%%%%%%%%%%%%%
%%%
%%% Schmidt Decomposition Formulas for Any Two-Qubit State
%%%
%%%%%%%%%%%%%%%%%%%%%%%%%%%%%%%%%%%%%%%%%%%%%%%%%%%
\section{Schmidt Decomposition Formulas for Two-Qubit pure States} \label{sec:1}

We derive explicit formulas for calculating the Schmidt decomposition of any two-qubit pure state. These formulas are obtained by following the existence proof of the Schmidt decomposition theorem~\cite{Schmidt1907,Nielsen2010,Wilde2013}, which is rooted in the existence of the singular value decomposition for matrices.

We consider a two-qubit system, denoted as $AB$, where the individual qubit subsystems are labeled $A$ and $B$. Each qubit resides in a two-dimensional Hilbert space, represented by the complex vector space $\mathbb{C}^2$. The computational basis for $\mathbb{C}^2$ consists of the basis vectors
\begin{equation}
\ket{0} \coloneqq \left[ \begin{array}{cc} 1 \\0 \end{array} \right], \quad
\ket{1} \coloneqq \left[ \begin{array}{cc} 0 \\1 \end{array} \right].
\end{equation}
For the two-qubit system, the computational basis is constructed as $\{\ket{jk}_{AB}\}$, where the basis vectors are defined by
\begin{equation}
\ket{jk}_{AB} \coloneqq \ket{j}_A \otimes \ket{k}_B, \label{eq:Combasis}
\end{equation}
and $\{\ket{j}_A\}$ and $\{\ket{k}_B\}$ are the computational bases for the qubit subsystems $A$ and $B$, respectively.

Let $\ket{\psi}_{AB}$ be an arbitrary pure state in the two-qubit system, expressed as 
\begin{equation}
\ket{\psi}_{AB} = \sum_{j=0}^1\sum_{k=0}^1 c_{jk} \ket{jk}_{AB} = \left[ \begin{array}{cc} c_{00} \\c_{01} \\c_{10} \\c_{11} \end{array} \right], \label{eq:Psi}
\end{equation}
where the coefficients $c_{jk}$ satisfy the normalization condition
\begin{equation}
\|\ket{\psi}\| = \sqrt{ \sum_{j=0}^1\sum_{k=0}^1 |c_{jk}|^2 } = 1.
\end{equation}
The computational basis provides a convenient framework for representing the pure state and serves as the foundation for deriving the Schmidt decomposition.

For a two-qubit pure state $\ket{\psi}_{AB}$ given in Eq.~(\ref{eq:Psi}), we define the associated $2 \times 2$ matrix $M_\psi$ as 
\begin{equation}
M_\psi \coloneqq \sum_{j=0}^1\sum_{k=0}^1 c_{jk} \ket{j}\bra{k} = \left[ \begin{array}{cc} c_{00} & c_{01} \\c_{10} & c_{11} \end{array} \right]. \label{eq:Matrix}
\end{equation}
The Gram matrix $G_\psi$~\cite{Horn1990,Bhatia2007}, corresponding to the same state, is defined as 
\begin{equation}
G_\psi \coloneqq M_\psi^\dagger M_\psi,
\end{equation}
where $M_\psi^\dagger$ is the conjugate transpose of $M_\psi$. The Gram matrix is Hermitian, positive semi-definite, and has trace 1. Therefore, its spectral decomposition~\cite{Wilde2013} can be expressed as 
\begin{equation}
G_\psi = \sum_{j=0}^1 r_j \ket{\varphi_j}\bra{\varphi_j},
\end{equation}
where $r_j$ are non-negative eigenvalues satisfying $r_0 + r_1 = 1$, and $\ket{\varphi_j}$ are the corresponding eigenstates.

The singular value decomposition~\cite{Trefethen1997,Strang2006} of $M_\psi$ is then given by 
\begin{equation}
M_\psi = \sum_{j=0}^1 \sqrt{r_j} \ket{\phi_j} \bra{\varphi_j}, \label{eq:SVD}
\end{equation}
where the states $\ket{\phi_j}$ are defined as 
\begin{equation}
\ket{\phi_j} \coloneqq \frac{1}{\sqrt{\bra{\varphi_j} G_\psi \ket{\varphi_j}}} M_\psi \ket{\varphi_j}.
\end{equation}
By comparing the matrix forms in Eqs.~(\ref{eq:Matrix}) and~(\ref{eq:SVD}), the coefficients $c_{jk}$ can be expressed as 
\begin{equation}
c_{jk} = \sum_{l=0}^1 \sqrt{r_l} \inn{j}{\phi_l} \inn{\varphi_l}{k}. 
\end{equation}
Consequently, the state $\ket{\psi}_{AB}$ is re-expressed as 
\begin{eqnarray} 
\ket{\psi}_{AB} 
&= \sum_{j=0}^1\sum_{k=0}^1 \left( \sum_{l=0}^1 \sqrt{r_l} \inn{j}{\phi_l} \inn{\varphi_l}{k} \right) \ket{jk} \\
&= \sum_{l=0}^1 \sqrt{r_l} \left( \sum_{j=0}^1 \ket{j} \inn{j}{\phi_l} \right) \otimes \left( \sum_{k=0}^1 \ket{k} \inn{k}{\varphi_l^*} \right) \\
&= \sum_{l=0}^1 \sqrt{r_l} \ket{\phi_l} \otimes \ket{\varphi_l^*}, \label{eq:SD}
\end{eqnarray}
where $\ket{\varphi_l^*}$ denotes the complex conjugate of $\ket{\varphi_l}$. This expression is the Schmidt decomposition of the state $\ket{\psi}_{AB}$, explicitly showing its entanglement structure in terms of the eigenvalues $r_l$ and the associated eigenstates.

From a practical perspective, the formula for calculating the Schmidt decomposition of any state $\ket{\psi}_{AB}$ depends on whether its coefficients satisfy the \emph{diagonal} condition 
\begin{equation}
c_{00}^* c_{01} + c_{10}^* c_{11} = 0, \label{eq:DiagonalCondition}
\end{equation}
where $c_{jk}^*$ denotes the complex conjugate of $c_{jk}$.

(i) If the diagonal condition in Eq.~(\ref{eq:DiagonalCondition}) is satisfied, the Gram matrix $G_\psi$ becomes diagonal. In this case, we denote the state $\ket{\psi}_{AB}$ as \emph{diagonal}, and the eigenvalues and eigenstates of $G_\psi$ are determined as 
\begin{eqnarray}
r_j = |c_{0j}|^2 + |c_{1j}|^2, \\
\ket{\varphi_j} = \ket{j},
\end{eqnarray}
with $j=0,1$.

Substituting these results into the Schmidt decomposition in Eq.~(\ref{eq:SD}) leads to the following proposition, which provides an explicit formula for diagonal states.

\begin{Prop}[Schmidt Decomposition: Diagonal] \label{prop:D}
Let $\ket{\psi}_{AB}$ be a pure state of the two-qubit system $AB$ as in Eq.~(\ref{eq:Psi}). 
When the state satisfies the diagonal condition in Eq.~(\ref{eq:DiagonalCondition}), its Schmidt decomposition is given by 
\begin{eqnarray}
\ket{\psi}_{AB} 
&= \sqrt{|c_{00}|^2 + |c_{10}|^2} \left( \frac{1}{\sqrt{|c_{00}|^2 + |c_{10}|^2}} \left[ \begin{array}{cc} c_{00} \\c_{10} \end{array} \right] \right) \otimes \ket{0} \nonumber \\
&\quad+\sqrt{|c_{01}|^2 + |c_{11}|^2} \left( \frac{1}{\sqrt{|c_{01}|^2 + |c_{11}|^2}} \left[ \begin{array}{cc} c_{01} \\c_{11} \end{array} \right] \right) \otimes \ket{1}. 
\end{eqnarray}
\end{Prop}

In Proposition~\ref{prop:D}, the Schmidt coefficients of the diagonal states are given by $\sqrt{|c_{0j}|^2 + |c_{1j}|^2}$ for $j=0,1$, and the states in parentheses represent the Schmidt basis in system $A$.

(ii) If the diagonal condition in Eq.~(\ref{eq:DiagonalCondition}) is not satisfied, the Gram matrix $G_\psi$ becomes non-diagonal, and the state $\ket{\psi}_{AB}$ is referred to as \emph{non-diagonal}. In this case, the eigenvalues and eigenstates of $G_\psi$ are determined as 
\begin{eqnarray}
r_j = \frac{1 + (-1)^j \sqrt{1 - 4|\det M_\psi|^2}}{2}, \\
\ket{y_j} = \left[ \begin{array}{cc} c_{00}^* c_{01} + c_{10}^* c_{11} \\r_j - |c_{00}|^2 - |c_{10}|^2 \end{array} \right], \label{eq:ORyj} \\
\ket{\varphi_j} = \frac{\ket{y_j}}{\| \ket{y_j} \|}.
\end{eqnarray}

The concurrence~\cite{Hill1997,Wootters1998}, a widely used measure of entanglement in two-qubit systems, quantifies the degree of entanglement in a two-qubit pure state. For a pure state $\ket{\psi}_{AB}$, the concurrence $C$ is given by 
\begin{equation}
C(\ket{\psi}) = 2 \left| c_{00}c_{11} - c_{01}c_{10} \right|, \label{eq:Concurrence}
\end{equation}
which vanishes, i.e., $C(\ket{\psi}) = 0$, if and only if the state is a product state.

These lead to the following proposition, which provides an explicit formula for the Schmidt decomposition of non-diagonal states.

\begin{Prop}[Schmidt Decomposition: Non-Diagonal] \label{prop:ND} 
Let $\ket{\psi}_{AB}$ be a pure state of the two-qubit system $AB$ as in Eq.~(\ref{eq:Psi}). 
When it does not satisfy the diagonal condition in Eq.~(\ref{eq:DiagonalCondition}), its Schmidt decomposition is expressed as 
\begin{equation}
\ket{\psi}_{AB} = \sum_{j=0}^1 \lambda_j \left( \frac{\ket{x_j}_A}{\|\ket{x_j}\|} \right) \otimes \left( \frac{\ket{y^*_j}_B}{\|\ket{y_j}\|} \right),
\end{equation}
where $\ket{y^*_j}$ denotes the complex conjugate of $\ket{y_j}$. The Schmidt coefficients $\lambda_j$ are given by 
\begin{equation}
\lambda_j \coloneqq \left( \frac{1 + (-1)^j \sqrt{1 - C(\ket{\psi})^2}}{2} \right)^{1/2}, 
\end{equation}
where the concurrence $C$ is presented in Eq.~(\ref{eq:Concurrence}), and the unnormalized vectors $\ket{y_j}$ and $\ket{x_j}$ are given by 
\begin{equation}
\ket{y_j} = \left[ \begin{array}{cc} c_{00}^* c_{01} + c_{10}^* c_{11} \\\lambda_j^2 - |c_{00}|^2 - |c_{10}|^2 \end{array} \right], \quad
\ket{x_j} = \left[ \begin{array}{cc} c_{00} & c_{01} \\c_{10} & c_{11} \end{array} \right] \ket{y_j}.
\end{equation}
\end{Prop}

Proposition~\ref{prop:ND} states that calculating the Schmidt decomposition of an arbitrary two-qubit pure state requires two distinct formulas. The appropriate formula depends on whether the Gram matrix of the state $\ket{\psi}_{AB}$ is diagonal or not.

In Proposition~\ref{prop:ND}, the Schmidt coefficients are expressed in terms of the concurrence rather than the coefficients $c_{ij}$ since the concurrence is given by the product of the two Schmidt coefficients, i.e., $C(\ket{\psi}) = 2\lambda_0\lambda_1$. This formulation provides a useful means to characterize entanglement in certain special cases. In particular, if a two-qubit pure state has zero concurrence, then one of the Schmidt coefficients must vanish, indicating that the state is a product state. Likewise, if the concurrence is equal to 1, the Schmidt coefficients are both $1/\sqrt{2}$, directly identifying the state as maximally entangled.

%%%%%%%%%%%%%%%%%%%%%%%%%%%%%%%%%%%%%%%%%%%%%%%%%%%
%%%
%%% Orthogonal Schmidt Decomposition Formulas for Two Orthogonal States
%%%
%%%%%%%%%%%%%%%%%%%%%%%%%%%%%%%%%%%%%%%%%%%%%%%%%%%
\section{Orthogonal Schmidt Decomposition Formulas for Two Orthogonal States} \label{sec:2}

In this section, we derive formulas for constructing arbitrary orthogonal two-qubit states in the form of Schmidt decomposition.

Since each two-qubit pure state can be classified as either a product state or an entangled state, we analyze four types of orthogonal state pairs: PP (both states are product), PE (the first state is product and the second is entangled), EP (the first state is entangled and the second is product), and EE (both states are entangled). In each type, the second state is determined based on the first state. For convenience, we use abbreviated notations to represent each pure state. For instance, in an EP-type pair, the first and second states are denoted as $\ket{\mathrm{Ep}}_{AB}$ and $\ket{\mathrm{eP}}_{AB}$, respectively. That is, uppercase letters indicate the order of the quantum states.

The type of a pair of two orthogonal states remains unchanged under local unitary transformations on subsystems $A$ and $B$. Conversely, this implies that once a formula for one specific pair of a given type is obtained, all pairs of that type can be constructed through local unitary transformations. Therefore, we present formulas for simplified pairs to streamline the derivation.

(i) We first present orthogonal Schmidt decomposition formulas for the PP-type pair, in which the two-qubit pure states are product states.

\begin{Prop}[PP] \label{prop:PP}
When one product state has the Schmidt decomposition 
\begin{equation}
\ket{\mathrm{Pp}}_{AB} = \ket{00}_{AB}, \label{eq:PP:Pp} 
\end{equation}
the other product state has one of the following Schmidt decompositions:
\begin{equation}
\ket{\mathrm{pP}}_{AB} = \ket{\alpha}_A \otimes \ket{1}_B, \label{eq:PP:pP1} 
\end{equation}
or 
\begin{equation}
\ket{\mathrm{pP}}_{AB} = \ket{1}_A \otimes \ket{\beta}_B, \label{eq:PP:pP2} 
\end{equation}
where $\ket{\alpha}_A$ and $\ket{\beta}_B$ are arbitrary pure single-qubit states.
\end{Prop}

Since both states are product states, they can be expressed as the tensor product of two single-qubit states. The first product state is set in its simplest form, as shown in Eq.~(\ref{eq:PP:Pp}). As any product state can be transformed into this form using local unitary transformations, this choice is made without loss of generality and helps simplify the analysis. For the two product states to be orthogonal, the single-qubit states corresponding to subsystems $A$ or $B$ must also be orthogonal. This establishes Proposition~\ref{prop:PP}.

(ii) We consider the PE-type pair, where the first state is a product state, and the second state, orthogonal to it, is an entangled state. The Schmidt decomposition of the second state is derived based on that of the first state. As discussed in Section~\ref{sec:1}, the Schmidt decomposition formula for an entangled state depends on whether the state is diagonal or non-diagonal.

\begin{Prop}[PE: Diagonal] \label{prop:PE:D}
Let a product state have the Schmidt decomposition 
\begin{equation}
\ket{\mathrm{Pe}}_{AB} = \ket{00}_{AB}. \label{eq:PE:Pe}
\end{equation}
The other state, which is entangled and diagonal, is characterized by two non-zero complex numbers $a$ and $b$ satisfying $|a|^2 + |b|^2 = 1$. Its Schmidt decomposition is then expressed as
\begin{equation}
\ket{\mathrm{pE}}_{AB} = |a| \left( \frac{a}{|a|} \ket{01}_{AB} \right) 
+ |b| \left( \frac{b}{|b|} \ket{10}_{AB} \right).
\end{equation}
\end{Prop}

When the first state has the product form given in Eq.~(\ref{eq:PE:Pe}), the second state, being entangled and orthogonal to the first, can be expressed as
\begin{equation}
a \ket{01}_{AB} + b \ket{10}_{AB} + c \ket{11}_{AB}, \label{eq:PE:pE:LC}
\end{equation}
where $a b \neq 0$. If $c = 0$, the second state becomes diagonal, and thus Proposition~\ref{prop:PE:D} is obtained by applying Proposition~\ref{prop:D} to the second state.

Otherwise, applying Proposition~\ref{prop:ND} leads to the following proposition.

\begin{Prop}[PE: Non-Diagonal] \label{prop:PE:ND}
Let a product state have the Schmidt decomposition in Eq.~(\ref{eq:PE:Pe}). 
The other state, which is entangled and non-diagonal, is characterized by three non-zero complex numbers $a$, $b$, and $c$ satisfying $|a|^2 + |b|^2 + |c|^2 = 1$. Its Schmidt decomposition is expressed as
\begin{equation}
\ket{\mathrm{pE}}_{AB} = \sum_{j=0}^1 \eta_j \left( \frac{\ket{x_j}_A}{\| \ket{x_j} \|} \right) \otimes \left( \frac{\ket{y^*_j}_B}{\| \ket{y_j} \|} \right),
\end{equation}
where the Schmidt coefficients $\eta_j$ are given by
\begin{equation}
\eta_j = \left( \frac{1 + (-1)^j \sqrt{1 - 4 |ab|^2}}{2} \right)^{1/2}, \label{eq:PE:pE:SC}
\end{equation}
and the orthogonal vectors $\ket{x_j}$ and $\ket{y_j}$ are defined as
\begin{equation}
\ket{x_j} = \left[ \begin{array}{cc} a \left( \eta_j^2 - |b|^2 \right) \\c \eta_j^2 \end{array} \right], \quad
\ket{y_j} = \left[ \begin{array}{cc} b^* c \\\eta_j^2 - |b|^2 \end{array} \right].
\end{equation}
\end{Prop}

(iii) We consider the EP type, where the first state is entangled and the second state is a product state. We present orthogonal Schmidt decomposition formulas for the EP type.

\begin{Prop}[EP] \label{prop:EP}
Let an entangled state have the Schmidt decomposition 
\begin{equation}
\ket{\mathrm{Ep}}_{AB} = \sqrt{\gamma} \ket{00}_{AB} + \sqrt{1-\gamma} \ket{11}_{AB}, \label{eq:EP:Ep}
\end{equation}
where $\gamma \in (0,1)$. A product state orthogonal to it is characterized by two complex numbers $a$ and $b$ satisfying
\begin{equation}
\left( |a| - \frac{\sqrt{\gamma}}{\sqrt{1-\gamma}} |b| \right) \left( |a| - \frac{\sqrt{1-\gamma}}{\sqrt{\gamma}} |b| \right) = 1.
\end{equation}
Its Schmidt decomposition is expressed as
\begin{eqnarray}
\ket{\mathrm{eP}}_{AB}
&= \left( \sqrt{a} \ket{0}_A \mp i \sqrt{\frac{\sqrt{\gamma}}{\sqrt{1-\gamma}}} \sqrt{b} \ket{1}_A \right) \nonumber \\
&\quad\otimes \left( \pm i \sqrt{\frac{\sqrt{1-\gamma}}{\sqrt{\gamma}}} \sqrt{b} \ket{0}_B + \sqrt{a} \ket{1}_B \right). \label{eq:EP:eP}
\end{eqnarray}
\end{Prop}

It is important to note that the formula in Eq.~(\ref{eq:EP:eP}) does not generally represent the Schmidt decomposition of the second state. To obtain the Schmidt decomposition, one must normalize its single-qubit vectors.

To derive the formula in Eq.~(\ref{eq:EP:eP}), we express the second state as a linear combination:
\begin{equation}
c \ket{00}_{AB} + a \ket{01}_{AB} + b \ket{10}_{AB} - \frac{\sqrt{\gamma}}{\sqrt{1-\gamma}} c \ket{11}_{AB}, \label{eq:EP:eP:LC}
\end{equation}
which is orthogonal to Eq.~(\ref{eq:EP:Ep}).
Since the second state is a product state, its concurrence must be zero. It follows that the coefficients satisfy
\begin{equation}
c = \pm i \sqrt{\frac{\sqrt{1-\gamma}}{\sqrt{\gamma}}} \sqrt{a b}.
\end{equation}
This establishes Proposition~\ref{prop:EP}.

(iv) Finally, we consider the EE-type pair, where both states are entangled. Depending on whether the second state is diagonal or non-diagonal, we present different formulas.

\begin{Prop}[EE: Diagonal] \label{prop:EE:D}
Let an entangled state have the Schmidt decomposition 
\begin{equation}
\ket{\mathrm{Ee}}_{AB} = \sqrt{\gamma} \ket{00}_{AB} + \sqrt{1-\gamma} \ket{11}_{AB}, \label{eq:EE:Ee}
\end{equation}
where $\gamma \in (0,1)$. The other entangled state, which is diagonal, is characterized by three complex numbers $a$, $b$, and $c$ satisfying the following three conditions:
\begin{eqnarray}
\frac{1}{1-\gamma} |a|^2 + |b|^2 + |c|^2 = 1, \label{eq:EE:Normal} \\
\sqrt{\gamma}a^2 + \sqrt{1-\gamma}bc \neq 0, \label{eq:EE:Entangled} \\
\sqrt{\gamma} a c^* - \sqrt{1-\gamma} a^* b = 0. \label{eq:EE:Diagonal}
\end{eqnarray}
Its Schmidt decomposition is then expressed as
\begin{equation}
\ket{\mathrm{eE}}_{AB}
=\zeta_0 \left( \frac{1}{\zeta_0} \left[ \begin{array}{cc} a \\c \end{array} \right] \right) \otimes \ket{0}
 +\zeta_1 \left( \frac{1}{\zeta_1} \left[ \begin{array}{cc} b \\- \frac{\sqrt{\gamma}}{\sqrt{1-\gamma}} a \end{array} \right] \right) \otimes \ket{1}, \label{eq:EE:eE}
\end{equation}
where the Schmidt coefficients $\zeta_j$ are defined as
\begin{equation}
\zeta_0 = \sqrt{|a|^2 + |c|^2}, \quad
\zeta_1 = \sqrt{|b|^2 + \frac{\gamma}{1-\gamma}|a|^2 }.
\end{equation}
\end{Prop}

To derive the formula for the second state, we express it as a linear combination in Eq.~(\ref{eq:EP:eP:LC}) so that the state is orthogonal to the first. The coefficients of this linear combination must ensure that the second state is unit, entangled, and diagonal, which correspond to the three conditions in Proposition~\ref{prop:EE:D}. Using the diagonal condition in Eq.~(\ref{eq:EE:Diagonal}) and Proposition~\ref{prop:D}, one can obtain the Schmidt decomposition formula in Eq.~(\ref{eq:EE:eE}).

When the second entangled state is non-diagonal, i.e., it does not satisfy the diagonal condition in Eq.~(\ref{eq:EE:Diagonal}), Proposition~\ref{prop:ND} leads to the following proposition.

\begin{Prop}[EE: Non-Diagonal] \label{prop:EE:ND}
Let an entangled state have the Schmidt decomposition in Eq.~(\ref{eq:EE:Ee}). 
The other entangled state, which is non-diagonal, is characterized by three complex numbers $a$, $b$, and $c$ satisfying the two conditions in Eqs.~(\ref{eq:EE:Normal}) and~(\ref{eq:EE:Entangled}), while not satisfying the condition in Eq.~(\ref{eq:EE:Diagonal}). 
Its Schmidt decomposition is expressed as
\begin{equation}
\ket{\mathrm{eE}}_{AB} = \sum_{j=0}^1 \delta_j \left( \frac{\ket{x_j}_A}{\| \ket{x_j} \|} \right) \otimes \left( \frac{\ket{y^*_j}_B}{\| \ket{y_j} \|} \right),
\end{equation}
where the Schmidt coefficients $\delta_j$ are given by
\begin{equation}
\delta_j = \left(1 + (-1)^j \sqrt{1 - 4 \left| \frac{\sqrt{\gamma}}{\sqrt{1-\gamma}}a^2 + bc \right|^2} \right)^{1/2} / \sqrt{2},
\end{equation}
and the unnormalized vectors $\ket{y_j}$ and $\ket{x_j}$ are defined as
\begin{equation}
\ket{y_j} = \left[ \begin{array}{cc} a^* b - \frac{\sqrt{\gamma}}{\sqrt{1-\gamma}} a c^* \\\delta_j^2 - |a|^2 - |c|^2 \end{array} \right], \quad
\ket{x_j} = \left[ \begin{array}{cc} a & b \\c & -\frac{\sqrt{\gamma}}{\sqrt{1-\gamma}} a \end{array} \right] \ket{y_j}.
\end{equation}
\end{Prop}

%%%%%%%%%%%%%%%%%%%%%%%%%%%%%%%%%%%%%%%%%%%%%%%%%%%
%%%
%%% Orthogonal Schmidt Decomposition Formulas for Three Mutually Orthogonal States
%%%
%%%%%%%%%%%%%%%%%%%%%%%%%%%%%%%%%%%%%%%%%%%%%%%%%%%
\section{Orthogonal Schmidt Decomposition Formulas for Three Mutually Orthogonal States} \label{sec:3}

We consider orthonormal sets of size 3. We classify all such sets into four types: PPP (all states are product), PPE (the first two states are product and the third is entangled), PEE (the first state is product and the remaining two are entangled), and EEE (all states are entangled).

In this section, we derive formulas for orthogonal Schmidt decompositions of the PPP and PPE types. However, for the PEE and EEE types, obtaining explicit formulas that characterize the orthogonality of three states is challenging. We explain the obstacle associated with deriving such formulas.

(i) We present the Schmidt decomposition formulas for three mutually orthogonal and product states.

\begin{Prop}[PPP] \label{prop:PPP}
When a product state is represented as
\begin{equation}
\ket{\mathrm{Ppp}}_{AB} = \ket{00}_{AB}, \label{eq:PPP:Ppp}
\end{equation}
the other product states are determined as
\begin{eqnarray}
\ket{\mathrm{pPp}}_{AB} &=& \ket{\alpha_0}_A \otimes \ket{1}_B, \label{eq:PPP:pPp1} \\
\ket{\mathrm{ppP}}_{AB} &=& \ket{\alpha_1}_A \otimes \ket{1}_B, \label{eq:PPP:ppP1}
\end{eqnarray}
or
\begin{eqnarray}
\ket{\mathrm{pPp}}_{AB} &=& \ket{1}_A \otimes \ket{\beta_0}_B, \\
\ket{\mathrm{ppP}}_{AB} &=& \ket{1}_A \otimes \ket{\beta_1}_B,
\end{eqnarray}
where $\{ \ket{\alpha_j}_A \}$ and $\{ \ket{\beta_j}_B \}$ denote arbitrary orthonormal bases of the qubit subsystems.
\end{Prop}

When the first product state $\ket{\mathrm{Ppp}}_{AB}$ has the Schmidt decomposition given in Eq.~(\ref{eq:PPP:Ppp}), the remaining states must be described by one of the Schmidt decompositions in Eqs.~(\ref{eq:PP:pP1}) and~(\ref{eq:PP:pP2}). Thus, the following four cases arise.

\begin{enumerate} 
\item[(1)] Consider the case where the remaining states share the same product form as in Eq.~(\ref{eq:PP:pP1}): 
\begin{eqnarray}
\ket{\mathrm{pPp}}_{AB} = \ket{\alpha}_A \otimes \ket{1}_B, \\
\ket{\mathrm{ppP}}_{AB} = \ket{\alpha'}_A \otimes \ket{1}_B, 
\end{eqnarray}
where $\ket{\alpha}_A$ and $\ket{\alpha'}_A$ are any single-qubit states for the qubit subsystem $A$. Since they are orthogonal, the single-qubit states form an orthonormal basis for subsystem $A$.

\item[(2)] When the product states $\ket{\mathrm{pPp}}_{AB}$ and $\ket{\mathrm{ppP}}_{AB}$ are represented in Eqs.~(\ref{eq:PP:pP1}) and~(\ref{eq:PP:pP2}), respectively, their respective single-qubit states $\ket{\alpha}_A$ and $\ket{\beta}_B$ must satisfy 
\begin{equation}
\inn{1}{\alpha}=0 \quad \mathrm{or} \quad \inn{1}{\beta}=0. 
\end{equation}
The single-qubit states can thus be represented as 
\begin{equation}
\ket{\alpha}_A=e^{i\theta}\ket{0}_A \quad \mathrm{or} \quad \ket{\beta}_B=e^{i\theta}\ket{0}_B, 
\end{equation}
where $\theta\in\mathbb{R}$.

\item[(3)] We consider the case where the second and third states are given by Eqs.~(\ref{eq:PP:pP2}) and~(\ref{eq:PP:pP1}), respectively. This scenario mirrors the second case, with the order of the quantum states interchanged.

\item[(4)] Finally, in the case where two product states are represented in the same product form as in Eq.~(\ref{eq:PP:pP2}), swapping the order of the qubit systems transforms it into the first case. Consequently, a symmetric result is obtained.
\end{enumerate}

By applying a local unitary transformation, the second and third cases can be respectively reduced to the first and last cases. This completes the proof of Proposition~\ref{prop:PPP}.

(ii) We derive the Schmidt decomposition formulas for three mutually orthogonal states, where the first two states are product states and the third state is entangled. Once the form of the first state is determined as 
\begin{equation} 
\ket{\mathrm{Ppe}}_{AB} = \ket{00}_{AB}, \label{eq:PPE1:Ppe} 
\end{equation} 
from Proposition~\ref{prop:PP}, the linear combination of the second product state can be classified into the following cases: 
\begin{eqnarray} 
\ket{\mathrm{pPe}}_{AB} = \ket{11}_{AB}, \label{eq:PPE1:pPe} \\ 
\ket{\mathrm{pPe}}_{AB} = a\ket{01}_{AB} + b\ket{11}_{AB}, \label{eq:PPE2:pPe} \\ 
\ket{\mathrm{pPe}}_{AB} = a\ket{10}_{AB} + b\ket{11}_{AB}, \label{eq:PPE3:pPe} 
\end{eqnarray} 
where $a$ and $b$ are non-zero complex numbers satisfying $|a|^2 + |b|^2 = 1$. Thus, we obtain the following three distinct cases.

\begin{Prop}[PPE: Case 1] \label{prop:PPE1} 
Let the first and second product states have the Schmidt decompositions given in Eqs.~(\ref{eq:PPE1:Ppe}) and~(\ref{eq:PPE1:pPe}), respectively. An entangled and diagonal state orthogonal to them can be characterized by two non-zero complex numbers, $c$ and $d$, satisfying $|c|^2 + |d|^2 = 1$. 
Its Schmidt decomposition can then be expressed as 
\begin{equation} 
\ket{\mathrm{ppE}}_{AB} = |c| \left( \frac{c}{|c|}\ket{01}_{AB} \right) + |d| \left( \frac{d}{|d|}\ket{10}_{AB} \right). \label{eq:PPE1:ppE} 
\end{equation} 
\end{Prop}

The first two states in Proposition~\ref{prop:PPE1} are in product form, as given in Eqs.~(\ref{eq:PPE1:Ppe}) and (\ref{eq:PPE1:pPe}), which implies that they already correspond to their Schmidt decomposition.

\begin{Prop}[PPE: Case 2] \label{prop:PPE2} 
Let the two product states have the Schmidt decompositions given in Eqs.~(\ref{eq:PPE1:Ppe}) and~(\ref{eq:PPE2:pPe}), respectively. 
An entangled state orthogonal to them is characterized by two non-zero complex numbers, $c$ and $d$, satisfying $|c|^2 + |d|^2 = 1$. 
Its Schmidt decomposition is given by 
\begin{equation} 
\ket{\mathrm{ppE}}_{AB} = \sum_{j=0}^1 \kappa_j \left( \frac{\ket{x_j}_A}{\| \ket{x_j} \|} \right) \otimes \left( \frac{\ket{y^*_j}_B}{\| \ket{y_j} \|} \right), \label{eq:PPE2:ppE} 
\end{equation} 
where the Schmidt coefficients $\kappa_j$ are given by 
\begin{equation} 
\kappa_j = \left( \frac{1 + (-1)^j \sqrt{1 - 4|bcd|^2}}{2} \right)^{1/2}, \label{eq:PPE2:ppE:SC} 
\end{equation} 
and the orthogonal vectors $\ket{x_j}$ and $\ket{y_j}$ are defined as 
\begin{equation} 
\ket{x_j} = \left[ \begin{array}{cc} b^*c(\kappa_j^2 - |d|^2) \\ -a^*c \kappa_j^2 \end{array} \right], \quad 
\ket{y_j} = \left[ \begin{array}{cc} -a^*cd^* \\ \kappa_j^2 - |d|^2 \end{array} \right]. 
\end{equation} 
\end{Prop}

For the product states in Proposition~\ref{prop:PPE2}, the entangled state can be expressed as the linear combination 
\begin{equation}
c\left( b^*\ket{01}_{AB} - a^* \ket{11}_{AB} \right) + d\ket{10}_{AB}, \label{eq:PPE2:ppE:LC}
\end{equation}
which is orthogonal to the first two states. By applying Proposition~\ref{prop:ND} to this entangled and non-diagonal state, the Schmidt decomposition in Eq.~(\ref{eq:PPE2:ppE}) is obtained.

\begin{Prop}[PPE: Case 3] \label{prop:PPE3} 
Let the two product states have the Schmidt decompositions in Eq.~(\ref{eq:PPE1:Ppe}) and~(\ref{eq:PPE3:pPe}), respectively.
An entangled state orthogonal to them can be characterized by two non-zero complex numbers $c$ and $d$, satisfying $|c|^2 + |d|^2 = 1$. 
Its Schmidt decomposition can then be expressed as 
\begin{equation}
\ket{\mathrm{ppE}}_{AB} = \sum_{j=0}^1 \nu_j \left( \frac{\ket{x_j}_A}{\| \ket{x_j} \|} \right) \otimes \left( \frac{\ket{y^*_j}_B}{\| \ket{y_j} \|} \right), \label{eq:PPE3:ppE} 
\end{equation}
where the Schmidt coefficients $\nu_j$ are given by 
\begin{equation}
\nu_j = \left( \frac{1 + (-1)^j \sqrt{1 - 4|bcd|^2}}{2} \right)^{1/2}, \label{eq:PPE3:ppE:SC} 
\end{equation}
and the orthogonal vectors $\ket{x_j}$ and $\ket{y_j}$ are defined as 
\begin{equation}
\ket{x_j} = \left[ \begin{array}{cc} c(\nu_j^2 - |bd|^2) \\-a^*d \nu_j^2 \end{array} \right], \quad 
\ket{y_j} = \left[ \begin{array}{cc} -a^*b|d|^2 \\\nu_j^2 - |bd|^2 \end{array} \right].
\end{equation}
\end{Prop}

Given the product states in Proposition~\ref{prop:PPE3}, the entangled state can be written as the linear combination 
\begin{equation}
c\ket{01}_{AB} + d\left( b^*\ket{10}_{AB} - a^* \ket{11}_{AB} \right), \label{eq:PPE3:ppE:LC} 
\end{equation}
which guarantees that the three states are mutually orthogonal and that the entangled state is non-diagonal. 
To obtain the Schmidt decomposition in Eq.~(\ref{eq:PPE3:ppE}), one applies Proposition~\ref{prop:ND} to this linear combination.

(iii) Unlike the PPP and PPE types, deriving explicit Schmidt decomposition formulas for the PEE and EEE types poses significant challenges.

Since quantum states in a two-qubit system are unit vectors and can be either product or entangled, constructing an orthonormal set requires careful consideration of both normalization and separability. In this work, we derive explicit Schmidt decomposition formulas that account for these conditions. Specifically, each quantum state is expressed as a linear combination, and the number of terms in this combination is reduced based on these conditions. The fewer terms required to represent a state, the simpler the resulting Schmidt decomposition formula becomes.

Among these conditions, separability plays the most significant role in reducing the number of coefficients. The number of product states determines the extent to which the coefficients in the linear combination can be simplified, making it easier to impose orthogonality. For the PPP and PPE types, where two or three product states exist, one can derive relatively simple Schmidt decomposition formulas. In contrast, for the PEE type, consider the case where the product state has the Schmidt decomposition 
\begin{equation}
\ket{\mathrm{Pee}}_{AB} = \ket{00}_{AB}. 
\end{equation}
The remaining entangled states can then be expressed as linear combinations 
\begin{eqnarray}
\ket{\mathrm{pEe}}_{AB} = a_0\ket{01}_{AB} + a_1\ket{10}_{AB} + a_2\ket{11}_{AB}, \\
\ket{\mathrm{peE}}_{AB} = b_0\ket{01}_{AB} + b_1\ket{10}_{AB} + b_2\ket{11}_{AB}. 
\end{eqnarray}
In this case, even if the normalization and orthogonality conditions reduce the number of coefficients, the resulting formulas become excessively complex and impractical. For similar reasons, deriving useful formulas for the EEE type is even more challenging, since there are no product states.

%%%%%%%%%%%%%%%%%%%%%%%%%%%%%%%%%%%%%%%%%%%%%%%%%%%
%%%
%%% Orthogonal Schmidt Decomposition Formulas for Orthonormal Bases
%%%
%%%%%%%%%%%%%%%%%%%%%%%%%%%%%%%%%%%%%%%%%%%%%%%%%%%
\section{Orthogonal Schmidt Decomposition Formulas for Orthonormal Bases} \label{sec:4}

We consider the orthonormal bases of a two-qubit system $AB$, where the basis vectors are either product states or entangled states. These orthonormal bases are classified into five types: PPPP (all states are product), PPPE (the first three states are product, and the fourth state is entangled), PPEE (the first two states are product, and the remaining states are entangled), PEEE (the first state is product, and the remaining states are entangled), and EEEE (all states are entangled).

In this section, we show that a PPPE-type orthonormal basis cannot exist. For the PPPP and PPEE types, we derive explicit Schmidt decomposition formulas. For the PEEE and EEEE types, we consider special cases where some of the entangled states are maximally entangled and derive explicit formulas for these cases.

%%%%%%%%%%%%%%%%%%%%%%%%%%%%%%%%%%%%%%%%%%%%%%%%%%%
%%% Non-Existence of PPPE-Type Orthonormal Bases
%%%%%%%%%%%%%%%%%%%%%%%%%%%%%%%%%%%%%%%%%%%%%%%%%%%
\subsection{Non-Existence of PPPE-Type Orthonormal Bases} \label{sec:PPPE}

An orthonormal basis of the PPPE type does not exist. To demonstrate this, we prove that if three basis vectors of an orthonormal basis are product, the remaining one must also be product.

Let four mutually orthogonal pure states be denoted as $\ket{\mathrm{Pppe}}_{AB}$, $\ket{\mathrm{pPpe}}_{AB}$, $\ket{\mathrm{ppPe}}_{AB}$, and $\ket{\mathrm{pppE}}_{AB}$, and assume that the first three states are product. The three product states form a PPP-type orthonormal set, and thus, according to Proposition~\ref{prop:PPP}, these states must be expressed in one of two possible forms.

When the first product state $\ket{\mathrm{Pppe}}_{AB}$ has the product form given in Eq.~(\ref{eq:PPP:Ppp}) and the other product states $\ket{\mathrm{pPpe}}_{AB}$ and $\ket{\mathrm{ppPe}}_{AB}$ are given in the product forms in Eqs.~(\ref{eq:PPP:pPp1}) and~(\ref{eq:PPP:ppP1}), a two-qubit orthonormal basis can be constructed with the following basis vectors: 
\begin{eqnarray}
\ket{\alpha_0}_A \otimes \ket{0}_B, \\
\ket{\alpha_0}_A \otimes \ket{1}_B, \\
\ket{\alpha_1}_A \otimes \ket{0}_B, \\
\ket{\alpha_1}_A \otimes \ket{1}_B. 
\end{eqnarray}
Since the last state $\ket{\mathrm{pppE}}_{AB}$ is orthogonal to both the second and third states, it must be expressed in terms of the constructed basis as 
\begin{equation} 
\left( a_0\ket{\alpha_0}_A + a_1\ket{\alpha_1}_A \right) \otimes \ket{0}_B. \label{eq:PPPE:pppE:LC} 
\end{equation} 
The same conclusion holds for the other case. Thus, when the first three states of an orthonormal basis are product, the final state must necessarily be a product state.

\begin{Thm}[PPPE: Impossible] \label{thm:PPPE} 
For any orthonormal basis of a two-qubit system, if three states in the basis are product, the remaining one must also be product. This implies that an orthonormal set of the PPPE type cannot exist in a two-qubit system. 
\end{Thm}

Theorem~\ref{thm:PPPE} can alternatively be proven in the framework of the unextendible product basis (UPB)~\cite{Bennett1999}. If a PPPE-type orthonormal basis existed, then its subset, the PPP-type orthonormal set, would form a UPB. According to Theorem 2 in Ref.~\cite{Bennett1999}, the members of a UPB cannot be perfectly distinguished using local positive operator-valued measurements and classical communication. However, it is well known that any set of orthogonal product states in a bipartite quantum system with dimensions $2 \times n$ is distinguishable using local measurements~\cite{Walgate2002}. This contradiction leads to the conclusion that a PPPE-type orthonormal basis cannot exist. Instead, our proof relies solely on orthogonality and does not use the concepts of UPBs or state discrimination.

%%%%%%%%%%%%%%%%%%%%%%%%%%%%%%%%%%%%%%%%%%%%%%%%%%%
%%% Orthogonal Schmidt Decomposition Formulas for PPPP-Type Orthonormal Bases
%%%%%%%%%%%%%%%%%%%%%%%%%%%%%%%%%%%%%%%%%%%%%%%%%%%
\subsection{Orthogonal Schmidt Decomposition Formulas for PPPP-Type Orthonormal Bases} \label{sec:PPPP}

We present the Schmidt decomposition formulas for PPPP-type orthonormal bases. By applying local unitary transformations to these formulas, one can construct arbitrary PPPP-type orthonormal bases.

\begin{Thm}[PPPP] \label{thm:PPPP} 
When a product state is given by 
\begin{equation}
\ket{\mathrm{Pppp}}_{AB} = \ket{00}_{AB}, \label{eq:PPPP:Pppp} 
\end{equation}
the remaining product states must take one of the following forms: 
\begin{eqnarray}
\ket{\mathrm{pPpp}}_{AB} = \ket{\alpha_0}_A \otimes \ket{1}_B, \\
\ket{\mathrm{ppPp}}_{AB} = \ket{\alpha_1}_A \otimes \ket{1}_B, \\
\ket{\mathrm{pppP}}_{AB} = \ket{10}_{AB}, 
\end{eqnarray}
or 
\begin{eqnarray}
\ket{\mathrm{pPpp}}_{AB} = \ket{1}_A \otimes \ket{\beta_0}_B, \\
\ket{\mathrm{ppPp}}_{AB} = \ket{1}_A \otimes \ket{\beta_1}_B, \\
\ket{\mathrm{pppP}}_{AB} = \ket{01}_{AB}, 
\end{eqnarray}
where $\{ \ket{\alpha_j}_A \}$ and $\{ \ket{\beta_j}_B \}$ denote orthonormal bases for the qubit subsystems. 
\end{Thm}

When the first product state is given by Eq.~(\ref{eq:PPPP:Pppp}), the second and third product states take one of two possible forms, as stated in Proposition~\ref{prop:PPP}. When the second and third states are represented as in Eqs.~(\ref{eq:PPP:pPp1}) and~(\ref{eq:PPP:ppP1}), the fourth state must be given by the linear combination in Eq.~(\ref{eq:PPPE:pppE:LC}) due to the orthogonality conditions. For the fourth state to be orthogonal to the first one, the single-qubit states on subsystem $A$ must satisfy 
\begin{equation}
\left( a_0^*\bra{\alpha_0} + a_1^*\bra{\alpha_1} \right) \ket{0} = 0. 
\end{equation}
Thus, the Schmidt decomposition of the fourth state is determined. Rearranging the order of the states and exchanging qubit subsystems $A$ and $B$ yields the same conclusion in the other case.

%%%%%%%%%%%%%%%%%%%%%%%%%%%%%%%%%%%%%%%%%%%%%%%%%%%
%%% Orthogonal Schmidt Decomposition Formulas for PPEE-Type Orthonormal Bases
%%%%%%%%%%%%%%%%%%%%%%%%%%%%%%%%%%%%%%%%%%%%%%%%%%%
\subsection{Orthogonal Schmidt Decomposition Formulas for PPEE-Type Orthonormal Bases} \label{sec:PPEE}

We derive the Schmidt decomposition formulas for PPEE-type orthonormal bases. Assume that the first product state has the Schmidt decomposition given in Eq.~(\ref{eq:PPE1:Ppe}). Then, the first three states of the PPEE-type basis form a PPE-type orthonormal set. Consequently, the Schmidt decomposition of the remaining entangled state is determined according to the three propositions in Section~\ref{sec:3}.

(i) In the first case, where the second and third states are described by Eqs.~(\ref{eq:PPE1:pPe}) and~(\ref{eq:PPE1:ppE}), the linear combination of the fourth state corresponds to its Schmidt decomposition.

\begin{Thm}[PPEE: Case 1] \label{thm:PPEE1} 
When the two product states and one entangled state of a PPEE-type orthonormal basis have the Schmidt decompositions 
\begin{eqnarray}
\ket{\mathrm{Ppee}}_{AB} = \ket{00}_{AB}, \label{eq:PPEE1:Ppee} \\
\ket{\mathrm{pPee}}_{AB} = \ket{11}_{AB}, \\
\ket{\mathrm{ppEe}}_{AB} = |a| \left( \frac{a}{|a|}\ket{01}_{AB} \right) + |b| \left( \frac{b}{|b|}\ket{10}_{AB} \right),
\end{eqnarray}
the Schmidt decomposition of the other entangled state is obtained as 
\begin{equation}
\ket{\mathrm{ppeE}}_{AB} = |b| \left( \frac{b^*}{|b|}\ket{01}_{AB} \right) + |a| \left( \frac{-a^*}{|a|}\ket{10}_{AB} \right).
\end{equation}
\end{Thm}

(ii) In the second case, where the second and third states of the PPEE-type orthonormal basis are given by Eqs.~(\ref{eq:PPE2:pPe}) and~(\ref{eq:PPE2:ppE:LC}), respectively, the orthogonality condition implies that the fourth state can be expressed as 
\begin{equation}
d^* ( b^*\ket{01}_{AB} - a^* \ket{11}_{AB} ) - c^*\ket{10}_{AB}, \label{eq:PPEE2:ppeE:LC} 
\end{equation}
which is non-diagonal. Thus, Proposition~\ref{prop:ND} leads to the following theorem.

\begin{Thm}[PPEE: Case 2] \label{thm:PPEE2} 
The first product state of a PPEE-type orthonormal basis is given by Eq.~(\ref{eq:PPEE1:Ppee}), and the other product state and one entangled state have the Schmidt decompositions 
\begin{eqnarray}
\ket{\mathrm{pPee}}_{AB} = a\ket{01}_{AB} + b\ket{11}_{AB}, \label{eq:PPEE2:pPee} \\
\ket{\mathrm{ppEe}}_{AB} = \sum_{j=0}^1 \kappa_j \left( \frac{\ket{x_j}_A}{\| \ket{x_j} \|} \right) \otimes \left( \frac{\ket{y^*_j}_B}{\| \ket{y_j} \|} \right),
\end{eqnarray}
where $a$ and $b$ are non-zero complex numbers such that $|a|^2 + |b|^2 = 1$, and the Schmidt coefficients and bases of the third state are presented in Proposition~\ref{prop:PPE2}. 
The Schmidt decomposition of the other entangled state is obtained as 
\begin{equation}
\ket{\mathrm{ppeE}}_{AB} 
= \sum_{j=0}^1 
\kappa_j \left( \frac{\ket{z_j}_A}{\| \ket{z_j} \|} \right) \otimes \left( \frac{\ket{y_{j\oplus1}^*}_B}{\| \ket{y_{j\oplus1}} \|} \right), \label{eq:PPEE2:ppeE} 
\end{equation}
where $ j \oplus 1 $ denotes the exclusive OR operation between $ j $ and 1, and the unnormalized vectors $\ket{z_j}_A$ are defined as 
\begin{equation}
\ket{z_j}_A = \left[ \begin{array}{cc} b^*\left(\kappa_j^2 - |c|^2\right) \\- a^*\kappa_j^2 \end{array} \right]. \label{eq:PPEE2:ppeE:A} 
\end{equation}
\end{Thm}

(iii) In the third case, where the second and third states are described by Eqs.~(\ref{eq:PPE3:pPe}) and~(\ref{eq:PPE3:ppE:LC}), the fourth state can be represented as 
\begin{equation}
d^*\ket{01}_{AB} - c^* ( b^*\ket{10}_{AB} - a^* \ket{11}_{AB} ). \label{eq:PPEE3:ppeE:LC} 
\end{equation}
In this case, the fourth state is non-diagonal, and Proposition~\ref{prop:ND} implies the following theorem.

\begin{Thm}[PPEE: Case 3] \label{thm:PPEE3} 
The first product state of a PPEE-type orthonormal basis is given by Eq.~(\ref{eq:PPEE1:Ppee}), and the other product state and one entangled state have the Schmidt decompositions 
\begin{eqnarray}
\ket{\mathrm{pPee}}_{AB} = a\ket{10}_{AB} + b\ket{11}_{AB}, \label{eq:PPEE3:pPee} \\
\ket{\mathrm{ppEe}}_{AB} = \sum_{j=0}^1 \nu_j \left( \frac{\ket{x_j}_A}{\| \ket{x_j} \|} \right) \otimes \left( \frac{\ket{y^*_j}_B}{\| \ket{y_j} \|} \right),
\end{eqnarray}
where $a$ and $b$ are non-zero complex numbers such that $|a|^2 + |b|^2 = 1$, and the Schmidt coefficients and bases of the third state are presented in Proposition~\ref{prop:PPE3}. 
The Schmidt decomposition of the other entangled state is obtained as 
\begin{equation}
\ket{\mathrm{ppeE}}_{AB} 
= \sum_{j=0}^1 
\nu_j \left( \frac{\ket{x_{j\oplus1}}_A}{\|\ket{x_{j\oplus1}}\|} \right) \otimes \left( \frac{\ket{w_j^*}_B}{\|\ket{w_j}\|} \right), \label{eq:PPEE3:ppeE} 
\end{equation}
where the vectors $\ket{w_j^*}$ are defined as 
\begin{equation}
\ket{w_j^*}_B = \left[ \begin{array}{cc} - ab^*|c|^2 \\\nu_j^2 - |bc|^2 \end{array} \right]. \label{eq:PPEE3:ppeE:B} 
\end{equation}
\end{Thm}

These propositions indicate that the two entangled states in a PPEE-type basis not only share the same concurrence but also have identical Schmidt bases for a single-qubit subsystem. Detailed proofs of Theorems~\ref{thm:PPEE2} and~\ref{thm:PPEE3} can be found in \ref{app:PPEE2} and \ref{app:PPEE3}, respectively.

%%%%%%%%%%%%%%%%%%%%%%%%%%%%%%%%%%%%%%%%%%%%%%%%%%%
%%% Special Case of the PEEE Type: PMEE Type
%%%%%%%%%%%%%%%%%%%%%%%%%%%%%%%%%%%%%%%%%%%%%%%%%%%
\subsection{Special Case of the PEEE Type: PMEE Type} \label{sec:PMEE}

Deriving explicit formulas for PEEE-type orthonormal bases is challenging because such bases contain only one product basis vector. In this subsection, we consider PMEE-type orthonormal bases, which form a special case of the PEEE type and include a single maximally entangled state. We derive the Schmidt decomposition formulas for PMEE-type bases.

Specifically, the PMEE-type orthonormal bases consist of one product state, one maximally entangled state, and two entangled states. Once the form of the product state is determined, the Schmidt decomposition of the maximally entangled state can be characterized by two real numbers.

\begin{Prop}[PM] \label{prop:PM} 
When a product state is given by 
\begin{equation}
\ket{\mathrm{Pm}}_{AB} = \ket{00}_{AB}, \label{eq:PM:Pm} 
\end{equation}
any maximally entangled state orthogonal to this product state can be expressed in the form of the Schmidt decomposition as 
\begin{equation}
\ket{\mathrm{pM}}_{AB} = \frac{1}{\sqrt{2}} \left( e^{i\theta} \ket{01}_{AB} + e^{i\theta'} \ket{10}_{AB} \right), \label{eq:PM:pM} 
\end{equation}
where $\theta$ and $\theta'$ are arbitrary real numbers. 
\end{Prop}

To prove Proposition~\ref{prop:PM}, consider an arbitrary entangled state orthogonal to the product state given in Eq.~(\ref{eq:PM:Pm}). 
According to Proposition~\ref{prop:PE:ND}, the Schmidt coefficients of the entangled state are given by Eq.~(\ref{eq:PE:pE:SC}). To ensure that its Schmidt coefficients are equal to $1/\sqrt{2}$, the following condition must be satisfied: 
\begin{equation}
4|ab|^2 = 1. 
\end{equation}
This implies that the coefficient $c$ of the linear combination in Eq.~(\ref{eq:PE:pE:LC}) becomes zero, meaning that the maximally entangled state is diagonal. 
Thus, by Proposition~\ref{prop:PE:D}, the Schmidt decomposition of the maximally entangled state is given in Eq.~(\ref{eq:PM:pM}).

The following theorem provides orthogonal Schmidt decomposition formulas for the PMEE-type orthonormal bases.

\begin{Thm}[PMEE] \label{thm:PMEE} 
Assume that the product state and the maximally entangled state of the PMEE-type orthonormal basis have the Schmidt decompositions 
\begin{eqnarray}
\ket{\mathrm{Pmee}}_{AB} = \ket{00}_{AB}, \label{eq:PMEE:Pmee} \\
\ket{\mathrm{pMee}}_{AB} = \frac{1}{\sqrt{2}} \left( e^{i\theta} \ket{01}_{AB} + e^{i\theta'} \ket{10}_{AB} \right), \label{eq:PMEE:pMee} 
\end{eqnarray}
where $\theta, \theta' \in \mathbb{R}$. The remaining entangled states are characterized by a real number $\theta''$ and a complex number $c$ with $0 < |c| < 1/\sqrt{2}$, and their Schmidt decompositions are given by 
\begin{eqnarray}
\ket{\mathrm{pmEe}}_{AB} = \sum_{j=0}^1 \xi_j \left( \frac{\ket{x_j}_A}{\| \ket{x_j} \|} \right) \otimes \left( \frac{\ket{y_j^*}_B}{\| \ket{y_j} \|} \right), \label{eq:PMEE:pmEe} \\
\ket{\mathrm{pmeE}}_{AB} = \sum_{j=0}^1 \upsilon_j \left( \frac{\ket{z_j}_A}{\| \ket{z_j} \|} \right) \otimes \left( \frac{\ket{w_j^*}_B}{\| \ket{w_j} \|} \right), \label{eq:PMEE:pmeE} 
\end{eqnarray}
where the Schmidt coefficients are given by 
\begin{eqnarray}
\xi_j = \left( \frac{1 + (-1)^j \sqrt{1 - 4 |c|^4}}{2} \right)^{1/2}, \label{eq:PMEE:pmEe:SC} \\
\upsilon_j = \left( \frac{1 + (-1)^j 2|c| \sqrt{1 - |c|^2}}{2} \right)^{1/2}, \label{eq:PMEE:pmeE:SC} 
\end{eqnarray}
and the related single-qubit vectors are given by 
\begin{eqnarray}
\ket{x_j} = \left[ \begin{array}{cc} (-1)^j c \\e^{i\theta''} \xi_j \end{array} \right], \label{eq:PMEE:mmEe:A} \\
\ket{y_j^*} = \left[ \begin{array}{cc} -e^{-i(\theta-\theta'+\theta'')} c \\ (-1)^j \xi_j \end{array} \right], \\
\ket{z_j} = \left[ \begin{array}{cc} (-1)^j e^{-i\theta''}\sqrt{\upsilon_0\upsilon_1}|c| \\-c^* \upsilon_j \end{array} \right], \label{eq:PMEE:mmeE:A} \\
\ket{w_j^*} = \left[ \begin{array}{cc} e^{-i(\theta-\theta'+\theta'')} \sqrt{\upsilon_0\upsilon_1} c \\ (-1)^j |c| \upsilon_j \end{array} \right]. 
\end{eqnarray}
\end{Thm}

Notably, the Schmidt basis vectors of the third state exhibit the property that their first components have magnitudes of $|c|$, while their second components have magnitudes of $\xi_j$. The same property also holds for the Schmidt basis vectors of the fourth state. The proof of Theorem~\ref{thm:PMEE} is presented in \ref{app:PMEE}.

%%%%%%%%%%%%%%%%%%%%%%%%%%%%%%%%%%%%%%%%%%%%%%%%%%%
%%% Special Case of the EEEE Type: MMEE Type
%%%%%%%%%%%%%%%%%%%%%%%%%%%%%%%%%%%%%%%%%%%%%%%%%%%
\subsection{Special Case of the EEEE Type: MMEE Type} \label{sec:MMEE}

Finding Schmidt decomposition formulas for EEEE-type orthonormal bases is more challenging than those of the PEEE type, since they do not include any product state. Instead, we consider a special case of the EEEE type, which is referred to as the MMEE type.

The MMEE-type orthonormal bases consist entirely of entangled states, including two maximally entangled states. The following theorems provide orthogonal Schmidt decomposition formulas for the MMEE type.

\begin{Thm}[MMEE: Diagonal] \label{thm:MMEE:D} 
Assume that the two maximally entangled states of the MMEE-type orthonormal basis have the Schmidt decompositions 
\begin{eqnarray}
\ket{\mathrm{Mmee}}_{AB} = \frac{1}{\sqrt{2}} \big( \ket{00}_{AB} + \ket{11}_{AB} \big), \label{eq:MMEE:Mmee} \\
\ket{\mathrm{mMee}}_{AB} = \frac{1}{\sqrt{2}} \left( e^{i\theta} \ket{01}_{AB} + e^{i\theta'} \ket{10}_{AB} \right), \label{eq:MMEE:mMee} 
\end{eqnarray}
where $\theta, \theta' \in \mathbb{R}$. 
The remaining states, which are diagonal, are characterized by two complex numbers $a$ and $b$ satisfying the following three conditions: 
\begin{eqnarray}
|a|^2 + |b|^2 = \frac{1}{2}, \label{eq:MMEE:Normal} \\
e^{i(\theta'-\theta)} b^2 \neq a^2, \label{eq:MMEE:Entangled} \\
D \coloneqq e^{i\frac{(\theta'-\theta)}{2}} a^*b + e^{-i\frac{(\theta'-\theta)}{2}}ab^* = 0, \label{eq:MMEE:Diagonal} 
\end{eqnarray}
and their Schmidt decompositions are given by 
\begin{eqnarray}
\ket{\mathrm{mmEe}}_{AB} 
&=& \frac{1}{\sqrt{2}} \left( \sqrt{2} \left[ \begin{array}{cc} a \\-e^{i(\theta'-\theta)}b \end{array} \right] \right)_A \otimes \ket{0}_B \nonumber \\
& &\quad+\frac{1}{\sqrt{2}} \left( \sqrt{2} \left[ \begin{array}{cc} b \\-a \end{array} \right] \right)_A \otimes \ket{1}_B, \\
\ket{\mathrm{mmeE}}_{AB} 
&=& \frac{1}{\sqrt{2}} \left( \sqrt{2} \left[ \begin{array}{cc} b^* \\e^{i(\theta'-\theta)}a^* \end{array} \right] \right)_A \otimes \ket{0}_B \nonumber \\
& &\quad+\frac{1}{\sqrt{2}} \left( \sqrt{2} \left[ \begin{array}{cc} -a^* \\-b^* \end{array} \right] \right)_A \otimes \ket{1}_B.
\end{eqnarray}
\end{Thm}

In Theorem~\ref{thm:MMEE:D}, not every maximally entangled state orthogonal to the one in Eq.~(\ref{eq:MMEE:Mmee}) can be represented in the Schmidt decomposition given in Eq.~(\ref{eq:MMEE:mMee}). When the maximally entangled states are given in the forms of Eqs.~(\ref{eq:MMEE:Mmee}) and~(\ref{eq:MMEE:mMee}), the third state $ \ket{\mathrm{mmEe}}_{AB} $, which is orthogonal to the first two states, can be expressed as the following linear combination: 
\begin{equation}
a\ket{00}_{AB} + b\ket{01}_{AB} - e^{i(\theta'-\theta)} b\ket{10}_{AB} - a\ket{11}_{AB}. \label{eq:MMEE:mmEe:LC} 
\end{equation}

Since the fourth state $ \ket{\mathrm{mmeE}}_{AB} $ is orthogonal to the first and second states, it must also take the same form as in Eq.~(\ref{eq:MMEE:mmEe:LC}). For the fourth state to be orthogonal to the third, it must be given by 
\begin{equation}
b^*\ket{00}_{AB} - a^*\ket{01}_{AB} + e^{i(\theta'-\theta)} a^*\ket{10}_{AB} - b^*\ket{11}_{AB}. \label{eq:MMEE:mmeE:LC} 
\end{equation}

The conditions in Eqs.~(\ref{eq:MMEE:Normal}) and~(\ref{eq:MMEE:Entangled}) are required to ensure that both the third and fourth states are normalized and entangled. They share the same diagonal condition, which is given in Eq.~(\ref{eq:MMEE:Diagonal}).

By applying the formula in Proposition~\ref{prop:D} to the linear combinations in Eqs.~(\ref{eq:MMEE:mmEe:LC}) and~(\ref{eq:MMEE:mmeE:LC}), we obtain the Schmidt decomposition formulas for the third and fourth entangled states in Theorem~\ref{thm:MMEE:D}. In this case, all entangled states in the MMEE-type orthonormal basis are maximally entangled.

When the third and fourth states do not satisfy the diagonal condition in Eq.~(\ref{eq:MMEE:Diagonal}), i.e., $D\neq0$, Proposition~\ref{prop:ND} leads to the following theorem.

\begin{Thm}[MMEE: Non-Diagonal] \label{thm:MMEE:ND} 
Assume that the two maximally entangled states are given by the forms in Eqs.~(\ref{eq:MMEE:Mmee}) and~(\ref{eq:MMEE:mMee}). The remaining entangled and non-diagonal states are characterized by two complex numbers $a$ and $b$ satisfying the two conditions in Eqs.~(\ref{eq:MMEE:Normal}) and~(\ref{eq:MMEE:Entangled}), but do not satisfy the diagonal condition in Eq.~(\ref{eq:MMEE:Diagonal}). Their Schmidt decompositions are then given by 
\begin{eqnarray}
\ket{\mathrm{mmEe}}_{AB} = \sum_{j=0}^1 \tau_j \ket{\alpha_j}_A \otimes \ket{\beta_j}_B, \\
\ket{\mathrm{mmeE}}_{AB} = \sum_{j=0}^1 \tau_j (-1)^j\ket{\beta_j^*}_A \otimes \ket{\alpha_j^*}_B ,
\end{eqnarray}
where the Schmidt coefficients $\tau_j$ are given by 
\begin{equation}
\tau_j = \left( \frac{1 + (-1)^j \sqrt{1 - 4 \left|a^2 - e^{i(\theta'-\theta)}b^2 \right|^2}}{2} \right)^{1/2}. \label{eq:MMEE:ND:mmEe:SC} 
\end{equation}
The corresponding single-qubit vectors are given by 
\begin{eqnarray}
\ket{\alpha_j}_A = \frac{1}{\sqrt{2}} \frac{C_j}{|C_j|} \left[ \begin{array}{cc} -e^{-i\frac{(\theta'-\theta)}{2}} \frac{D}{|D|} \\ (-1)^j \end{array} \right], \\
\ket{\beta_j}_B = \frac{1}{\sqrt{2}}\left[ \begin{array}{cc} e^{i\frac{(\theta'-\theta)}{2}} \frac{D}{|D|} \\ (-1)^j \end{array} \right],
\end{eqnarray}
where the coefficient $D$ is presented in Eq.~(\ref{eq:MMEE:Diagonal}), and $C_j$ are defined as 
\begin{equation}
C_j = e^{-i\frac{(\theta'-\theta)}{2}} \frac{D}{|D|} a + (-1)^j b. \label{eq:MMEE:ND:Cj} 
\end{equation}
\end{Thm} 

These theorems show that the last two entangled states in the MMEE-type orthonormal basis always have the same Schmidt coefficient. Detailed explanations of Theorems~\ref{thm:MMEE:D} and~\ref{thm:MMEE:ND} are provided in Appendices~D and~E, respectively.

%%%%%%%%%%%%%%%%%%%%%%%%%%%%%%%%%%%%%%%%%%%%%%%%%%%
%%%
%%% Examples
%%%
%%%%%%%%%%%%%%%%%%%%%%%%%%%%%%%%%%%%%%%%%%%%%%%%%%%
\section{Examples} \label{sec:Examples}

In this section, we present examples that apply our formulas to representative states and verify the resulting decompositions. We then explain how these results can be used to design state-preparation circuits for current quantum processors, without committing to any specific hardware platform.

Propositions~\ref{prop:D} and~\ref{prop:ND} are used to compute the Schmidt decomposition of a given two-qubit pure state. Consider the following two states:
\begin{eqnarray}
\ket{\xi_1}_{AB} &=& \sqrt{\frac{1}{3}} \ket{00}_{AB} + \sqrt{\frac{1}{6}} \ket{01}_{AB} + \sqrt{\frac{1}{6}} \ket{10}_{AB} - \sqrt{\frac{1}{3}} \ket{11}_{AB}, \\
\ket{\xi_2}_{AB} &=& \sqrt{\frac{1}{3}} \ket{00}_{AB} + \sqrt{\frac{1}{3}} \ket{01}_{AB} + \sqrt{\frac{1}{6}} \ket{10}_{AB} - \sqrt{\frac{1}{6}} \ket{11}_{AB},
\end{eqnarray}
where the first state satisfies the diagonal condition in Eq.~(\ref{eq:DiagonalCondition}) and is therefore diagonal, whereas the second state is non-diagonal. By applying Proposition~\ref{prop:D} to $\ket{\xi_1}_{AB}$, we obtain the following Schmidt decomposition:
\begin{equation}
\ket{\xi_1}_{AB} = \frac{1}{\sqrt{2}} \left[ \begin{array}{cc} \frac{\sqrt{2}}{\sqrt{3}} \\ \frac{1}{\sqrt{3}} \end{array} \right] \otimes \ket{0}
+ \frac{1}{\sqrt{2}} \left[ \begin{array}{cc} \frac{1}{\sqrt{3}} \\ -\frac{\sqrt{2}}{\sqrt{3}} \end{array} \right] \otimes \ket{1}.
\end{equation}
If, instead, one applies the non-diagonal formula to this state, the vectors appearing in Eq.~(\ref{eq:ORyj}) both become the zero vector, and thus this does not yield a Schmidt decomposition. Next, by applying Proposition~\ref{prop:ND} to the second state $\ket{\xi_2}_{AB}$, we obtain
\begin{equation}
\ket{\xi_2}_{AB} = \sqrt{\frac{2}{3}} \ket{0} \otimes \ket{+} + \sqrt{\frac{1}{3}} \ket{1} \otimes \ket{-}.
\end{equation}
In this case as well, if one applies the formula for the diagonal case to the non-diagonal state $\ket{\xi_2}_{AB}$, one arrives at the following invalid decomposition:
\begin{equation}
\frac{1}{\sqrt{2}} \left[ \begin{array}{cc} \frac{\sqrt{2}}{\sqrt{3}} \\ \frac{1}{\sqrt{3}} \end{array} \right] \otimes \ket{0}
+ \frac{1}{\sqrt{2}} \left[ \begin{array}{cc} \frac{\sqrt{2}}{\sqrt{3}} \\ -\frac{1}{\sqrt{3}} \end{array} \right] \otimes \ket{1},
\end{equation}
which does not yield the correct Schmidt coefficients and whose single-qubit vectors on system $A$ are not orthogonal. Taken together, these two examples demonstrate why two distinct Schmidt decomposition formulas are necessary.

We now consider the computational basis vectors in Eq.~(\ref{eq:Combasis}) together with the following Bell basis vectors~\cite{Bennett1993}:
\begin{eqnarray}
\ket{\Phi^\pm}_{AB} &=& \frac{1}{\sqrt{2}}\big(\ket{00}_{AB} \pm \ket{11}_{AB}\big), \\
\ket{\Psi^\pm}_{AB} &=& \frac{1}{\sqrt{2}}\big(\ket{01}_{AB} \pm \ket{10}_{AB}\big).
\end{eqnarray}
These are well-known representatives of PPPP-type and EEEE-type orthonormal bases, respectively. Except for Propositions~\ref{prop:D} and~\ref{prop:ND}, the remaining results are used to construct orthonormal sets.

As a first illustration, consider the task of finding a product state orthogonal to $\ket{00}_{AB}$. Proposition~\ref{prop:PP} gives the general form of all such product states. For the entangled case, the general form of states orthogonal to $\ket{00}_{AB}$ is determined by Propositions~\ref{prop:PE:D} and~\ref{prop:PE:ND}. Proposition~\ref{prop:PE:D} covers the case in which the computational basis serves as the Schmidt basis, whereas Proposition~\ref{prop:PE:ND} addresses the remaining non-diagonal cases. In particular, Proposition~\ref{prop:PE:ND} allows one to obtain entangled states orthogonal to $\ket{00}_{AB}$ that are not restricted to the Bell states $\ket{\Psi^\pm}_{AB}$. For example, by setting $a=1/\sqrt{2}$ and $b=c=1/2$ in that proposition, one obtains the following entangled state $\ket{\xi_3}_{AB}$, written explicitly in Schmidt form and orthogonal to $\ket{00}_{AB}$:
\begin{eqnarray}
\ket{\xi_3}_{AB} &=\sqrt{\frac{1}{2}+\frac{1}{2\sqrt{2}}}\ket{+}\otimes
\left(
\frac{1}{2\sqrt{1+\frac{1}{\sqrt{2}}}}
\left[
\begin{array}{cc}
1 \\
1+\sqrt{2}
\end{array}
\right]
\right) \nonumber \\
&\quad+\sqrt{\frac{1}{2}-\frac{1}{2\sqrt{2}}}\big(-\ket{-}\big)\otimes
\left(
\frac{1}{2\sqrt{1-\frac{1}{\sqrt{2}}}}
\left[
\begin{array}{cc}
1 \\
1-\sqrt{2}
\end{array}
\right]
\right). \label{eq:examplePE}
\end{eqnarray}
This example demonstrates that our construction yields orthogonal entangled states beyond the Bell family and makes their Schmidt data explicit.

Moreover, our results also address settings with larger orthogonal families. For instance, by combining the computational basis vectors with the Bell basis vectors, one can see that a PPPE-type orthonormal basis cannot be obtained in that manner. This observation alone, however, does not establish that PPPE-type bases are impossible in general. Theorem~\ref{thm:PPPE} provides a rigorous impossibility result, thereby clarifying this point.

Our results can also be used to construct two-qubit mixed states. Any mixed state admits a spectral decomposition, i.e., it is a convex combination of mutually orthogonal two-qubit pure states. As a simple instance, suppose one wishes to build a rank-2 mixed state $\rho_{AB}$ that includes the product state $\ket{00}_{AB}$. Using Propositions~\ref{prop:PP},~\ref{prop:PE:D}, and~\ref{prop:PE:ND}, one can construct the desired second eigenstate. Moreover, since our constructions are presented in Schmidt form, the single-qubit reductions are immediate in this case. Specifically, among the states orthogonal to $\ket{00}_{AB}$, we consider the entangled state $\ket{\xi_3}_{AB}$ defined in Eq.~(\ref{eq:examplePE}); let the weights $\omega_j>0$ with $\omega_0+\omega_1=1$ be given. Then the two-qubit mixed state is
\begin{equation}
\rho_{AB} = \omega_0 \ket{00}\bra{00} + \omega_1 \ket{\xi_3}\bra{\xi_3},
\end{equation}
and the reduced state on system $A$ is
\begin{equation}
\rho_A = \omega_0 \ket{0}\bra{0} + \omega_1\left(\frac{1}{2\sqrt{2}}
\left[
\begin{array}{cc}
\sqrt{2} & 1 \\
1 & \sqrt{2}
\end{array}
\right] \right).
\end{equation}
Accordingly, the reduced density matrix can be computed more readily by invoking our results.

Finally, beyond theoretical illustrations, we outline how our results apply to actual physical systems. A basic example is state preparation~\cite{Zylberman2024,Huggins2025,Long2025}. Contemporary quantum processors are implemented on diverse physical platforms—including superconducting circuits, trapped ions, photonic systems, and neutral atoms~\cite{Devoret2013,Blatt2012,OBrien2007,Saffman2010}—yet their operation is governed by the same principles of quantum mechanics; hence our results are platform independent. Concretely, consider a user who wishes to prepare a specific three-qubit pure state on a quantum device. A useful approach is to analyze the state’s entanglement structure across the $A|BC$ bipartition via its Schmidt decomposition, whose rank is either $1$ or $2$. If the rank is $1$, the target is separable across $A|BC$ and the task reduces to preparing a two-qubit state. In this case, one can obtain the Schmidt decomposition of the two-qubit state using Propositions~\ref{prop:D} and~\ref{prop:ND}. If the rank is $2$, the $BC$ component is a superposition of two mutually orthogonal two-qubit pure states, so our results for the PP, PE, EP, and EE types apply directly. Since the entanglement structure of the target three-qubit state is invariant under local unitaries on qubit systems, our formulas can serve to inform an intuitive circuit design.

%%%%%%%%%%%%%%%%%%%%%%%%%%%%%%%%%%%%%%%%%%%%%%%%%%%
%%%
%%% Conclusion
%%%
%%%%%%%%%%%%%%%%%%%%%%%%%%%%%%%%%%%%%%%%%%%%%%%%%%%
\section{Conclusion} \label{sec:Conclusion}

In this work, we presented Schmidt decomposition formulas for mutually orthogonal two-qubit states. In Section~\ref{sec:1}, we classified two-qubit pure states under the diagonal condition and presented formulas for computing their Schmidt decompositions. 
In Section~\ref{sec:2}, we divided orthonormal sets of size 2 into the PP, PE, EP, and EE types and derived the orthogonal Schmidt decomposition formulas for all types of mutually orthogonal two-qubit states. 
Section~\ref{sec:3} extended this approach to three mutually orthogonal states. We categorized such orthonormal sets into the PPP, PPE, PEE, and EEE types and derived the Schmidt decomposition formulas for the PPP and PPE types. For the PEE and EEE types, we discussed the challenges in deriving analytical formulas. 
In Section~\ref{sec:4}, we classified orthonormal bases into the PPPP, PPPE, PPEE, PEEE, and EEEE types. We proved that PPPE-type orthonormal bases cannot exist. For the PPPP and PPEE types, we derived explicit orthogonal Schmidt decomposition formulas. For the PEEE and EEEE types, we analyzed special cases involving maximally entangled states---namely, the PMEE and MMEE types---and established the corresponding formulas. Section~\ref{sec:Examples} presented representative examples that verified our formulas and illustrated their use in designing state-preparation circuits.

Beyond these theoretical contributions, our formulas offer a practical framework for constructing two-qubit mixed states. Since any mixed state can be expressed in terms of its spectral decomposition, our results enable an explicit description of the eigenstate combinations that constitute this decomposition. In particular, by applying local unitary transformations to our formulas in Section~\ref{sec:2}, arbitrary rank-2 mixed states can be constructed.

Furthermore, our findings contribute to a deeper understanding of two-qubit mixed states. For instance, Theorem~\ref{thm:PPPE} demonstrates that, in order to construct a full-rank mixed state via spectral decomposition, the number of mutually orthogonal entangled states must be either zero or at least two. This highlights fundamental constraints in the construction of orthonormal bases in two-qubit systems.

Overall, our results address fundamental questions regarding the structure of orthonormal sets in two-qubit systems. We have shown that orthogonal Schmidt decomposition formulas provide a systematic approach to constructing orthonormal sets and bases. We hope that our findings will serve as a foundation for applications beyond the construction of two-qubit mixed states, extending to broader areas of quantum information science.

\ack
We are grateful to Donghoon Ha and Soojoon Lee for their valuable assistance and insightful discussions.
This study was supported by 2023 Research Grant from Kangwon National University.
The present Research has been conducted by the Research Grant of Kwangwoon University in 2025. 
This research was supported by Basic Science Research Program through the National Research Foundation of Korea (NRF) funded by the Ministry of Education (Grant No. NRF-2020R1I1A1A01058364, Grant No. RS-2023-00243988, and Grant No. RS-2025-25415913). This work was also supported by Institute of Information \& Communications Technology Planning \& Evaluation (IITP) grant funded by the Korea government (MSIT) (RS-2024-00437284).

\appendix

%%%%%%%%%%%%%%%%%%%%%%%%%%%%%%%%%%%%%%%%%%%%%%%%%%%%%%%%%%%%%%%%%%%%%%%%%%%%%%%%%%%%%%%
%%
%% Proof of Theorem~\ref{thm:PPEE2}
%%
%%%%%%%%%%%%%%%%%%%%%%%%%%%%%%%%%%%%%%%%%%%%%%%%%%%%%%%%%%%%%%%%%%%%%%%%%%%%%%%%%%%%%%%
\section{Proof of Theorem~\ref{thm:PPEE2}} \label{app:PPEE2}

We derive the Schmidt decomposition of the fourth state in Eq.~(\ref{eq:PPEE2:ppeE}). Assume that the first three states of the PPEE-type orthonormal basis have Schmidt decompositions given by Eqs.~(\ref{eq:PPEE1:Ppee}),~(\ref{eq:PPEE2:pPee}), and~(\ref{eq:PPE2:ppE:LC}), respectively. Then, the fourth state must be expressed as the linear combination in Eq.~(\ref{eq:PPEE2:ppeE:LC}), which corresponds to a non-diagonal entangled state. From Proposition~\ref{prop:ND}, its Schmidt decomposition is given by 
\begin{equation}
\ket{\mathrm{ppeE}}_{AB} = \sum_{j=0}^1 \mu_j \left( \frac{\ket{v_j}_A}{\|\ket{v_j}\|} \right) \otimes \left( \frac{\ket{w^*_j}_B}{\|\ket{w_j}\|} \right). \label{eq:PPEE2:ppeE:app} 
\end{equation}

Note that the concurrences of the third and fourth states are identical, and thus, they have identical Schmidt coefficients, i.e., $\mu_j = \kappa_j$, where $\kappa_j$ is given in Eq.~(\ref{eq:PPE2:ppE:SC}). According to Proposition~\ref{prop:ND}, the unnormalized states $\ket{v_j}_A$ and $\ket{w_j}_B$ can be expressed as 
\begin{eqnarray}
\ket{v_j}_A = - d^* 
\left[ \begin{array}{cc} b^*\left(\kappa_j^2 - |c|^2\right) \\- a^*\kappa_j^2 \end{array} \right]= - d^* \ket{z_j}_A, \\
\ket{w_j}_B = \left[ \begin{array}{cc} a^*cd^* \\\kappa_j^2 - |c|^2 \end{array} \right], 
\end{eqnarray}
where the vector $\ket{z_j}_A$ is given in Eq.~(\ref{eq:PPEE2:ppeE:A}). 
By using the identities $\kappa_0^2 + \kappa_1^2 =1$ and $|c|^2 + |d|^2=1$, it follows that 
\begin{equation}
\ket{w_j}_B = - \ket{y_{j\oplus1}}_B, 
\end{equation}
where the operation $j \oplus 1$ denotes the exclusive OR between the index $j$ and 1.

Consequently, the Schmidt decomposition of the fourth state in Eq.~(\ref{eq:PPEE2:ppeE:app}) can be rewritten as 
\begin{eqnarray}
\ket{\mathrm{ppeE}}_{AB}
&= \sum_{j=0}^1 \kappa_j \left( \frac{- d^* \ket{z_j}_A}{\|- d^* \ket{z_j}_A\|} \right) \otimes \left( \frac{- \ket{y_{j\oplus1}}_B}{\| \ket{y^*_{j\oplus1}} \|} \right) \\
&= \frac{d^*}{|d|} \sum_{j=0}^1 \kappa_j \left( \frac{\ket{z_j}_A}{\|\ket{z_j}\|} \right) \otimes \left( \frac{ \ket{y^*_{j\oplus1}}_B}{\| \ket{y_{j\oplus1}} \|} \right). 
\end{eqnarray}
Up to a global phase, this is equivalent to the Schmidt decomposition given in Eq.~(\ref{eq:PPEE2:ppeE}). 
Moreover, the global phases of the basis vectors do not affect their orthogonality; thus, they are omitted in our Schmidt decomposition formulas. This completes the proof of Theorem~\ref{thm:PPEE2}.

%%%%%%%%%%%%%%%%%%%%%%%%%%%%%%%%%%%%%%%%%%%%%%%%%%%%%%%%%%%%%%%%%%%%%%%%%%%%%%%%%%%%%%%
%%
%% Proof of Theorem~\ref{thm:PPEE3}
%%
%%%%%%%%%%%%%%%%%%%%%%%%%%%%%%%%%%%%%%%%%%%%%%%%%%%%%%%%%%%%%%%%%%%%%%%%%%%%%%%%%%%%%%%
\section{Proof of Theorem~\ref{thm:PPEE3}} \label{app:PPEE3}

We derive the Schmidt decomposition of the entangled state $\ket{\mathrm{ppeE}}_{AB}$ given in Eq.~(\ref{eq:PPEE3:ppeE}). Consider the case where the first three states of a PPEE-type orthonormal basis are given by the Schmidt decompositions in Eqs.~(\ref{eq:PPEE1:Ppee}),~(\ref{eq:PPEE3:pPee}), and~(\ref{eq:PPE3:ppE:LC}), respectively. Then the fourth state becomes a non-diagonal and entangled state, and it is given by the linear combination in Eq.~(\ref{eq:PPEE3:ppeE:LC}). According to Proposition~\ref{prop:ND}, the Schmidt decomposition of the fourth state is given by 
\begin{equation}
\ket{\mathrm{ppeE}}_{AB} = \sum_{j=0}^1 \xi_j \left( \frac{\ket{z_j}_A}{\|\ket{z_j}\|} \right) \otimes \left( \frac{\ket{w^*_j}_B}{\|\ket{w_j}\|} \right). \label{eq:PPEE3:ppeE:app} 
\end{equation}

In this case, the last two states share the same Schmidt coefficients, i.e., $ \xi_j = \nu_j $, where $ \nu_j $ is given in Eq.~(\ref{eq:PPE3:ppE:SC}). According to Proposition~\ref{prop:ND}, one can calculate the unnormalized state $\ket{z_j}_A$ as 
\begin{equation}
\ket{z_j}_A = \left[ \begin{array}{cc} d^*(\nu_j^2 - |bc|^2) \\a^*c^*\nu_j^2 \end{array} \right], 
\end{equation}
and the state $\ket{w^*_j}_B$ is given in Eq.~(\ref{eq:PPEE3:ppeE:B}).

Given the two identities, $\nu_0^2 + \nu_1^2 = 1$ and $|c|^2+|d|^2=1$, we obtain the identity
\begin{eqnarray}
&\quad\quad
\nu_j^2 \nu_k^2 = \left| bcd \right|^2 \\
&\Rightarrow
\left( |c|^2+|d|^2 \right) \nu_j^2 \nu_k^2 = \left| bcd \right|^2 \left( \nu_j^2 + \nu_k^2 \right) \\
&\Rightarrow
|c|^2 \nu_k^2 \left( \nu_j^2 - \left| bd \right|^2 \right) = |d|^2 \nu_j^2 \left( \left| bc \right|^2 - \nu_k^2 \right) \\
&\Rightarrow
\frac{c \left( \nu_j^2 - \left| bd \right|^2 \right)}{d \nu_j^2} = -\frac{d^* \left( \nu_k^2 - \left| bc \right|^2 \right)}{c^* \nu_k^2}
\end{eqnarray}
where the indices $ j $ and $ k $ are distinct, i.e., $ j \neq k $.
This leads to the following expression:
\begin{equation}
\ket{z_j}_A = -\frac{c^* \nu_j^2}{d \nu_{j\oplus1}^2}\ket{x_{j\oplus1}}_A.
\end{equation}

Consequently, the Schmidt decomposition of the fourth state in Eq.~(\ref{eq:PPEE3:ppeE:app}) can be rewritten as
\begin{eqnarray}
\ket{\mathrm{ppeE}}_{AB}
&= \sum_{j=0}^1 \nu_j \left( \frac{-\frac{c^* \nu_j^2}{d \nu_{j\oplus1}^2}\ket{x_{j\oplus1}}_A}{\left\| \frac{c^* \nu_j^2}{d \nu_{j\oplus1}^2}\ket{x_{j\oplus1}}_A \right\|} \right) \otimes \left( \frac{\ket{w^*_j}_B}{\|\ket{w_j}\|} \right) \\
&= -\frac{c^*|d|}{d|c^*|} \sum_{j=0}^1 \nu_j \left( \frac{\ket{x_{j\oplus1}}_A}{\left\| \ket{x_{j\oplus1}}_A \right\|} \right) \otimes \left( \frac{\ket{w^*_j}_B}{\|\ket{w_j}\|} \right).
\end{eqnarray}
Up to a global phase, this Schmidt decomposition is equivalent to the one given in Eq.~(\ref{eq:PPEE3:ppeE}). This completes the proof of Theorem~\ref{thm:PPEE3}.

%%%%%%%%%%%%%%%%%%%%%%%%%%%%%%%%%%%%%%%%%%%%%%%%%%%%%%%%%%%%%%%%%%%%%%%%%%%%%%%%%%%%%%%
%%
%% Proof of Theorem~\ref{thm:PMEE}
%%
%%%%%%%%%%%%%%%%%%%%%%%%%%%%%%%%%%%%%%%%%%%%%%%%%%%%%%%%%%%%%%%%%%%%%%%%%%%%%%%%%%%%%%%
\section{Proof of Theorem~\ref{thm:PMEE}} \label{app:PMEE}

When the product state and the maximally entangled state of the PMEE-type orthonormal basis are given by the Schmidt decompositions in Eqs.~(\ref{eq:PMEE:Pmee}) and~(\ref{eq:PMEE:pMee}), we obtain the linear combinations of the remaining two entangled states and then derive their Schmidt decomposition formulas.

(i) We determine the linear combination of the third state. Since this state is orthogonal to the product state, it can be expressed as 
\begin{equation}
a_1 \ket{01}_{AB} + a_2 \ket{10}_{AB} + a_3 \ket{11}_{AB}, 
\end{equation}
where the coefficients $ a_j $ satisfy the normalization condition and the constraints $ a_1 \neq 0 $ and $ a_2 \neq 0 $. Moreover, since this state is orthogonal to the maximally entangled state, the coefficient $ a_2 $ is given by 
\begin{equation}
a_2 = - e^{i(\theta'-\theta)}a_1, 
\end{equation}
where $ \theta' \in \mathbb{R} $. For the state to be a unit vector, the coefficient $a_3$ must be 
\begin{equation}
a_3 = e^{i\theta''} \sqrt{1-2|a_1|^2}, 
\end{equation}
with $0 \le |a_1| \le 1/\sqrt{2}$ and $\theta'' \in \mathbb{R}$.

Observe that if $ |a_1| = 0 $ or $ |a_1| = 1/\sqrt{2} $, the PMEE-type orthonormal basis reduces to a PPEE-type orthonormal basis. Thus, the third state $\ket{\mathrm{pmEe}}_{AB}$ can be represented as 
\begin{equation}
c \ket{01}_{AB} - e^{i(\theta'-\theta)}c \ket{10}_{AB} + e^{i\theta''} \sqrt{1-2|c|^2} \ket{11}_{AB}, \label{eq:PMEE:pmEe:LC:app} 
\end{equation}
where $0<|c|< 1/\sqrt{2}$ and $\theta''\in\mathbb{R}$. It follows that the third state is non-diagonal.

(ii) By applying Proposition~\ref{prop:ND} to the linear combination in Eq.~(\ref{eq:PMEE:pmEe:LC:app}), we obtain the Schmidt decomposition of the third state, given in Eq.~(\ref{eq:PMEE:pmEe}), where the Schmidt coefficients $\xi_j$ are given in Eq.~(\ref{eq:PMEE:pmEe:SC}). The unnormalized vector $\ket{y_j}_B$ in the Schmidt decomposition is calculated as 
\begin{equation}
\ket{y_j}_B = \left[ \begin{array}{cc} -e^{i(\theta-\theta'+\theta'')} c^* \sqrt{1-2|c|^2} \\\xi_j^2 - |c|^2 \end{array} \right]. 
\end{equation}

The eigenvalues $\xi_j^2$ of the Gram matrix for the third state satisfy the following equation: 
\begin{equation}
\xi_j^4 - \xi_j^2 + |c|^4 = 0. 
\end{equation}
It follows that the following identities hold: 
\begin{eqnarray}
\left( \xi_j^2 - |c|^2 \right)^2 = \xi_j^2 \left( 1 - 2|c|^2 \right), \\
\xi_j^2 - |c|^2 = (-1)^j \xi_j \sqrt{ 1 - 2|c|^2 }, \label{eq:PMEE:Identity:app} 
\end{eqnarray}
and the vector $\ket{y_j}_B$ can be rewritten as 
\begin{equation}
\ket{y_j}_B = \sqrt{1-2|c|^2} \left[ \begin{array}{cc} -e^{i(\theta-\theta'+\theta'')} c^* \\ (-1)^j \xi_j \end{array} \right]. 
\end{equation}

By normalizing this vector, the positive scalar $\sqrt{1-2|c|^2}$ cancels out, and thus we can redefine $\ket{y_j}_B$ as 
\begin{equation}
\ket{y_j}_B 
= \left[ \begin{array}{cc} -e^{i(\theta-\theta'+\theta'')} c^* \\ (-1)^j \xi_j \end{array} \right]. 
\end{equation}
Then, using the identity in Eq.~(\ref{eq:PMEE:Identity:app}), the other unnormalized vector $\ket{x_j}_A$ is given by 
\begin{equation}
\ket{x_j}_A = \xi_j\left[ \begin{array}{cc} (-1)^j c \\e^{i\theta''} \xi_j \end{array} \right]. 
\end{equation}
For consistency, we remove the positive scalar $\xi_j$ and redefine $\ket{x_j}_A$ as in Eq.~(\ref{eq:PMEE:mmEe:A}).

(iii) Similar to the third state, the fourth state is orthogonal to the first two states. Thus, it can be expressed as the following linear combination: 
\begin{equation}
b \ket{01}_{AB} - e^{i(\theta'-\theta)}b \ket{10}_{AB} + e^{i\theta'''}\sqrt{1-2|b|^2} \ket{11}_{AB}, 
\end{equation}
where $0<|b|< 1/\sqrt{2}$ and $\theta''' \in \mathbb{R}$. Imposing the orthogonality condition with the third state yields 
\begin{equation}
b = - e^{i(\theta'''-\theta'')}\frac{c}{|c|}\sqrt{\frac{1}{2}-|c|^2}. 
\end{equation}
Removing the global phase, the fourth state can be expressed as 
\begin{eqnarray}
\ket{\mathrm{pmeE}}_{AB} 
&= e^{-i\theta''}\sqrt{\frac{1}{2}-|c|^2} \ket{01}_{AB} \nonumber \\ 
&\quad - e^{i(\theta'-\theta-\theta'')}\sqrt{\frac{1}{2}-|c|^2} \ket{10}_{AB} \nonumber \\ 
&\quad - \sqrt{2}c^* \ket{11}_{AB}, \label{eq:PMEE:pmeE:LC:app} 
\end{eqnarray}
which shows that it is a non-diagonal state.

(iv) Applying Proposition~\ref{prop:ND} to the linear combination in Eq.~(\ref{eq:PMEE:pmeE:LC:app}) yields the Schmidt decomposition of the fourth state, given in Eq.~(\ref{eq:PMEE:pmeE}), with the Schmidt coefficients $\upsilon_j$ given in Eq.~(\ref{eq:PMEE:pmeE:SC}).

The eigenvalues $\upsilon_j^2$ of the Gram matrix for the fourth state satisfy 
\begin{equation}
\upsilon_j^4 - \upsilon_j^2 + \left( \frac{1}{2}-|c|^2 \right)^2 = 0, 
\end{equation}
which leads to the following identities: 
\begin{eqnarray}
\left( \upsilon_j^2 - \left( \frac{1}{2}-|c|^2 \right) \right)^2 = 2 |c|^2 \upsilon_j^2, \\
\upsilon_j^2 - \left( \frac{1}{2}-|c|^2 \right) = (-1)^j \sqrt{2} |c| \upsilon_j. \label{eq:PMEE:Identity2:app} 
\end{eqnarray}
The unnormalized vector $\ket{w_j}_B$ is given by 
\begin{equation}
\ket{w_j}_B 
= \left[ \begin{array}{cc} e^{i(\theta-\theta'+\theta'')} \sqrt{1-2|c|^2} c^* \\\upsilon_j^2 - \left( \frac{1}{2}-|c|^2 \right) \end{array} \right]. 
\end{equation}
By simplifying the second component using Eq.~(\ref{eq:PMEE:Identity2:app}), we redefine the vector $\ket{w_j}_B$ as 
\begin{equation}
\ket{w_j}_B = 
\left[ \begin{array}{cc} e^{i(\theta-\theta'+\theta'')} \sqrt{\upsilon_0\upsilon_1} c^* \\ (-1)^j |c| \upsilon_j \end{array} \right], 
\end{equation}
where $\upsilon_0\upsilon_1=\frac{1}{2}-|c|^2$.
The other unnormalized vector $\ket{z_j}_A$ is given by 
\begin{eqnarray}
\ket{z_j}_A 
&= \left[ \begin{array}{cc} (-1)^j e^{-i\theta''}\sqrt{\frac{1}{2}-|c|^2}|c|\upsilon_j \\-c^* \left( \frac{1}{2}-|c|^2+(-1)^j\sqrt{2}|c|\upsilon_j \right) \end{array} \right] \\
&= \left[ \begin{array}{cc} (-1)^j e^{-i\theta''}\sqrt{\frac{1}{2}-|c|^2}|c| \upsilon_j \\-c^* \upsilon_j^2 \end{array} \right] \\
&= \upsilon_j \left[ \begin{array}{cc} (-1)^j e^{-i\theta''}\sqrt{\frac{1}{2}-|c|^2}|c| \\-c^* \upsilon_j \end{array} \right]. 
\end{eqnarray}
When normalizing this vector, the positive scalar $\upsilon_j$ cancels out, and we therefore redefine the state $\ket{z_j}_A$ as in Eq.~(\ref{eq:PMEE:mmeE:A}).

%%%%%%%%%%%%%%%%%%%%%%%%%%%%%%%%%%%%%%%%%%%%%%%%%%%%%%%%%%%%%%%%%%%%%%%%%%%%%%%%%%%%%%%
%%
%% Proof of Theorem~\ref{thm:MMEE:D}
%%
%%%%%%%%%%%%%%%%%%%%%%%%%%%%%%%%%%%%%%%%%%%%%%%%%%%%%%%%%%%%%%%%%%%%%%%%%%%%%%%%%%%%%%%
\section{Proof of Theorem~\ref{thm:MMEE:D}} \label{app:MMEED}

To prove Theorem~\ref{thm:MMEE:D}, we first derive the linear combinations of the third and fourth states. The first and second states of the MMEE-type orthonormal basis are given by the Schmidt decompositions in Eqs.~(\ref{eq:MMEE:Mmee}) and~(\ref{eq:MMEE:mMee}), respectively. In the computational basis of each qubit subsystem, the third state can be expressed as the linear combination 
\begin{equation}
a\ket{00}_{AB} + b\ket{01}_{AB} + c\ket{10}_{AB} + d\ket{11}_{AB}. 
\end{equation}
Since it is orthogonal to the first two states, its linear combination is given by Eq.~(\ref{eq:MMEE:mmEe:LC}).

Similarly, the fourth state, which is also orthogonal to the first two states, can be expressed as 
\begin{equation}
d_0\ket{00}_{AB} + d_1\ket{01}_{AB} - e^{i(\theta'-\theta)} d_1\ket{10}_{AB} - d_0\ket{11}_{AB}. 
\end{equation}
Furthermore, for the third and fourth states to be orthogonal, the coefficients $d_j$ must satisfy
\begin{equation}
b^*d_1 = -a^*d_0. 
\end{equation}
As scalar multiplication does not affect orthogonality between the third and fourth states, multiplying by $b^*$ results in the linear combination in Eq.~(\ref{eq:MMEE:mmeE:LC}).

By applying Proposition~\ref{prop:D} to the linear combinations in Eqs.~(\ref{eq:MMEE:mmEe:LC}) and~(\ref{eq:MMEE:mmeE:LC}), we derive the Schmidt decomposition formulas given in Theorem~\ref{thm:MMEE:D}.

%%%%%%%%%%%%%%%%%%%%%%%%%%%%%%%%%%%%%%%%%%%%%%%%%%%%%%%%%%%%%%%%%%%%%%%%%%%%%%%%%%%%%%%
%%
%% Proof of Theorem~\ref{thm:MMEE:ND}
%%
%%%%%%%%%%%%%%%%%%%%%%%%%%%%%%%%%%%%%%%%%%%%%%%%%%%%%%%%%%%%%%%%%%%%%%%%%%%%%%%%%%%%%%%
\section{Proof of Theorem~\ref{thm:MMEE:ND}} \label{app:MMEEND}

To prove Theorem~\ref{thm:MMEE:ND}, we begin with the linear combinations of the third and fourth states in Eqs.~(\ref{eq:MMEE:mmEe:LC}) and~(\ref{eq:MMEE:mmeE:LC}). Theorem~\ref{thm:MMEE:ND} addresses the case in which the third and fourth states are non-diagonal, i.e., $D \neq 0$, where $D$ is defined in Eq.~(\ref{eq:MMEE:Diagonal}).

(i) Applying Proposition~\ref{prop:ND} to Eq.~(\ref{eq:MMEE:mmEe:LC}), we obtain the Schmidt decomposition 
\begin{equation} 
\ket{\mathrm{mmEe}}_{AB} = \sum_{j=0}^1 \tau_j \left( \frac{\ket{x_j}_A}{\| \ket{x_j} \|} \right) \otimes \left( \frac{\ket{y_j^*}_B}{\| \ket{y_j} \|} \right), 
\end{equation} 
where the Schmidt coefficients $ \tau_j $ are given in Eq.~(\ref{eq:MMEE:ND:mmEe:SC}). According to the formula in Proposition~\ref{prop:ND}, the Schmidt basis for system $B$ is expressed as 
\begin{eqnarray} 
\ket{y_j}_B 
&= \left[ \begin{array}{cc} a^*b + e^{-i(\theta'-\theta)}ab^* \\ \tau_j^2 - |a|^2 - |b|^2 \end{array} \right] \\ 
&= \left[ \begin{array}{cc} a^*b + e^{-i(\theta'-\theta)}ab^* \\ \tau_j^2 - \frac{1}{2} \end{array} \right] \\ 
&= \left[ \begin{array}{cc} a^*b + e^{-i(\theta'-\theta)}ab^* \\ (-1)^j \left| D \right| \end{array} \right] \\ 
&= \left[ \begin{array}{cc} e^{-i\frac{(\theta'-\theta)}{2}} D \\ (-1)^j \left| D \right| \end{array} \right], 
\end{eqnarray} 
where the second equality follows from the condition in Eq.~(\ref{eq:MMEE:Normal}). In the third equality, the second component is simplified as 
\begin{equation} 
\tau_j^2 - \frac{1}{2} 
= (-1)^j \sqrt{\frac{1}{4} - \left|a^2 - e^{i(\theta'-\theta)}b^2 \right|^2} 
= (-1)^j \left| D \right|. 
\end{equation} 
By the definition of $D$, the first component is also simplified in the last equality. 
After normalization, the basis vectors are redefined as 
\begin{equation} 
\ket{y_j}_B = \frac{1}{\sqrt{2}}\left[ \begin{array}{cc} e^{-i\frac{(\theta'-\theta)}{2}} \frac{D}{|D|} \\ (-1)^j \end{array} \right]. 
\end{equation}
From this, the Schmidt basis for system $A$ is given by 
\begin{equation} 
\ket{x_j}_A = \frac{1}{\sqrt{2}} \left[ \begin{array}{cc} e^{-i\frac{(\theta'-\theta)}{2}} \frac{D}{|D|} a + (-1)^j b \\-e^{i\frac{(\theta'-\theta)}{2}} \frac{D}{|D|} b - (-1)^j a \end{array} \right]. 
\end{equation} 
Then, the basis vector can be rewritten as 
\begin{eqnarray} 
\ket{x_j}_A
&= \frac{1}{\sqrt{2}} C_j \left[ \begin{array}{cc} 1 \\-e^{i\frac{(\theta'-\theta)}{2}} \frac{D}{|D|} (-1)^j \end{array} \right] \\ 
&= \left( -e^{i\frac{(\theta'-\theta)}{2}} \frac{D}{|D|} \right) 
\frac{1}{\sqrt{2}} C_j \left[ \begin{array}{cc} -e^{-i\frac{(\theta'-\theta)}{2}} \frac{D}{|D|} \\ (-1)^j \end{array} \right],
\end{eqnarray}
where the coefficient $C_j$ is defined in Eq.~(\ref{eq:MMEE:ND:Cj}).
Since unit complex numbers independent of the index $j$ do not affect the orthogonality of two-qubit pure states, they can be omitted. Thus, after normalization, we obtain 
\begin{equation} 
\ket{x_j}_A 
= \frac{1}{\sqrt{2}} \frac{C_j}{|C_j|} \left[ \begin{array}{cc} -e^{-i\frac{(\theta'-\theta)}{2}} \frac{D}{|D|} \\ (-1)^j \end{array} \right]. 
\end{equation} 
Therefore, the Schmidt decomposition of the third entangled state $\ket{\mathrm{mmEe}}$ is expressed as 
\begin{equation} 
\sum_{j=0}^1 \tau_j 
\left( \frac{1}{\sqrt{2}} \frac{C_j}{|C_j|} \left[ \begin{array}{cc} -e^{-i\frac{(\theta'-\theta)}{2}} \frac{D}{|D|} \\ (-1)^j \end{array} \right] \right) 
\otimes 
\left( \frac{1}{\sqrt{2}}\left[ \begin{array}{cc} e^{i\frac{(\theta'-\theta)}{2}} \frac{D}{|D|} \\ (-1)^j \end{array} \right] \right). 
\end{equation}

(ii) Observe that the linear combination of $\ket{\mathrm{mmeE}}$ in Eq.~(\ref{eq:MMEE:mmeE:LC}) can be obtained by replacing $a$ and $b$ with $b^*$ and $-a^*$, respectively, in the linear combination of $\ket{\mathrm{mmEe}}$ in Eq.~(\ref{eq:MMEE:mmEe:LC}). This implies that the Schmidt decomposition of $\ket{\mathrm{mmeE}}$ can be directly derived from that of $\ket{\mathrm{mmEe}}$. Specifically, each variable undergoes the following substitutions: 
\begin{eqnarray} 
a &\longrightarrow& b^*, \\ 
b &\longrightarrow& -a^*, \\ 
\tau_j &\longrightarrow& \tau_j, \\ 
D &\longrightarrow& -D, \\ 
C_j &\longrightarrow& -\left( e^{-i\frac{(\theta'-\theta)}{2}} \frac{D}{|D|} (-1)^j \right)C_j^*. 
\end{eqnarray} 
As a result, the Schmidt decomposition of the fourth entangled state $\ket{\mathrm{mmeE}}$ is given by 
\begin{equation}
\sum_{j=0}^1 \tau_j (-1)^j 
\left( \frac{1}{\sqrt{2}}\left[ \begin{array}{cc} e^{-i\frac{(\theta'-\theta)}{2}} \frac{D}{|D|} \\ (-1)^j \end{array} \right] \right)
\otimes 
\left( \frac{1}{\sqrt{2}} \frac{C_j^*}{|C_j|} \left[ \begin{array}{cc} -e^{i\frac{(\theta'-\theta)}{2}} \frac{D}{|D|} \\ (-1)^j \end{array} \right] \right).
\end{equation}
This completes the proof of Theorem~\ref{thm:MMEE:ND}.

\section*{References}
\bibliography{TwoQubitSchDecom}

\end{document}